\documentclass{aa} 
\usepackage{graphicx}
\usepackage{txfonts}
\usepackage{xcolor}

\usepackage{natbib}
\usepackage{multirow}
\bibpunct{(}{)}{;}{a}{}{,} 

\begin{document}

\title{ Evolution of binary black holes in AGN accretion discs: \\
Disc-binary interaction and gravitational wave emission}
   
   \subtitle{}
 
   \author{W. Ishibashi 
          \inst{1}
          \and
          M. Gröbner
           \inst{2}
          }

 \institute{Physik-Institut, Universität Zürich, Zürich, Switzerland \label{inst1}
 \and Department of Physics, ETH Zürich, Zürich, Switzerland \label{inst2} \\
 \email{wako.ishibashi@physik.uzh.ch}
    }
    
       \date{Received; accepted}

\abstract{ Binary black hole (BBH) mergers are the primary sources of gravitational wave (GW) events detected by LIGO/Virgo. Binary black holes embedded in the accretion discs of active galactic nuclei (AGN) are possible candidates for such GW events. We develop an idealised analytic model for the orbital evolution of BBHs in AGN accretion discs by combining the evolution equations of disc-binary interaction and GW inspiral. We investigate the coupled `disc+GW'-driven evolution of BBHs transitioning from the disc-driven regime at large orbital separations into the GW-driven regime at small separations. In this evolution channel, BBH mergers are accelerated by a combination of orbital decay and orbital eccentricity growth in the disc-dominated regime. We provide a quantification of the resulting merger timescale $\tau_\text{merger}$, and analyse its dependence on both the accretion disc and binary orbital parameters. By computing the evolution of the orbital eccentricity as a function of the GW frequency, we predict that most binaries in AGN discs should have significant residual eccentricities ($e \sim 0.01-0.1$), potentially detectable by LISA. We further discuss the potentials and caveats of this particular BBH-in-AGN channel in the framework of binary evolutionary paths. }


   \authorrunning{ W. Ishibashi $\&$ M. Gröbner}
   \titlerunning{Evolution of BBHs in AGN accretion discs}

   \maketitle



\section{Introduction} 

Binary black holes (BBH) are one of the main cosmic sources of gravitational waves (GW) in the Universe. Most GW events detected so far by LIGO/Virgo arise from the merger of two stellar-mass black holes \citep{Abbott_et_2019_runs}. Different scenarios for the formation of such BBH systems have been proposed and discussed in the literature. The two main binary black hole formation channels are isolated evolution in galactic fields and dynamical evolution in dense stellar environments. In the case of isolated evolution, two massive stars in a binary system exchange mass through Roche lobe overflow (possibly evolving through a common envelope phase) and eventually collapse into black holes (BHs), giving rise to a BBH system \citep[e.g.][]{Belczynski_et_2016, Mandel_Farmer_2018}. In the dynamical formation scenario, the two BHs are independently formed by the collapse of individual massive stars (the two BHs do not form in the same binary), and only later pair up in dense stellar environments, such as globular clusters or nuclear star clusters. Due to mass segregation, the BHs tend to sink towards the centre of the stellar cluster, where they may form binaries by hardening through different three-body interactions \citep[e.g.][]{Mapelli_2018}. 

The predicted merger rates overlap for the two scenarios, and both channels can yield BBHs with a broad range of masses, consistent with those observed. Thus merger rates and BH masses are not enough to distinguish between the two formation paths. 
On the other hand, different BH spin and orbital eccentricity distributions are predicted for the two channels \citep{Rodriguez_et_2016, Breivik_et_2016, Dorazio_Samsing_2018}. In the case of isolated evolution, the individual BH spins should be aligned with the orbital angular momentum, and the binary is expected to have a negligible eccentricity. In contrast, in dense stellar environments, the BH spins and orbital angular momentum tend to be randomly oriented, and the binary is likely to have a non-negligible eccentricity. More precisely, field binaries are expected to have nearly negligible eccentricities in the Laser Interferometer Space Antenna (LISA) frequency band, whereas binaries formed in dense star clusters can have significant residual eccentricities (we recall that most binaries will have near-zero eccentricities in the LIGO band). 

Binary black holes are eventually driven to merger by gravitational wave emission. However the inspiral due to GW radiation only becomes relevant at small enough orbital separations. On large galactic scales, stellar scatterings can help harden the binary, but become inefficient on smaller scales. As a consequence, the binary may stall before GW inspiral can take over. This suggests the existence of an intermediate region where additional processes are required to bridge the gap. It has been argued that viscous torques in a dense gaseous environment may assist the binary orbital decay and thus accelerate the BBH merger \citep{Armitage_Natarajan_2005, Cuadra_et_2009}. In this picture, binaries embedded in a gaseous disc experience angular momentum loss to the surrounding gas (through the disc-binary interaction) and may be brought to the point where GW emission becomes dominant. 

An interesting possibility is the formation of BBHs within the accretion disc surrounding the super-massive black hole (SMBH) at the centre of active galactic nuclei (AGN) \citep[][]{Stone_et_2017, Bartos_et_2017, McKernan_et_2018, Tagawa_et_2019}. In the nuclear regions, a heavy concentration of stellar populations is expected, with stars originating either from the nuclear star cluster or formed in-situ in the outer self-gravitating parts of the accretion disc \citep[e.g.][]{Levin_2007}.  A fraction of such stellar remnants may end up as BBHs ground down into the AGN accretion disc. 
Binary black hole systems evolving in this particular gaseous environment could be driven towards merger through disc-binary interactions. However the actual efficiency of such gas-fostered inspiral is still an unsettled issue. In previous works, embedded binaries were usually assumed to merge within the AGN disc lifetime, on timescales comparable to the Salpeter time $t_\textrm{AGN} \sim 10^7$ yr \citep{Bartos_et_2017, Stone_et_2017, McKernan_et_2018}.

In this paper we develop an idealised analytic model for the evolution of BBHs in AGN accretion discs by combining disc-driven and GW-driven evolution equations (the coupled `disc+GW'-driven evolution). We find that a combination of the orbital decay and eccentricity growth in the disc-driven regime can facilitate the transition into the GW-driven regime and accelerate the BBH merger.  
We then quantify the merger timescale resulting from the coupled disc+GW-driven evolution, and we compute the evolution of the orbital eccentricity throughout the LISA frequency band. We further discuss the astrophysical implications of this particular BBH-in-AGN evolution channel, also recalling its limitations and outlining possible directions for future studies. 


\section{Binary black holes in active galactic nuclei accretion discs}
\label{Sect_BBHinAGN}

We set up a simple analytic model for the orbital evolution of BBHs embedded in the gaseous disc of an AGN. We rederive the basic equations governing the disc-binary interaction, with some insight provided by numerical simulations. A self-contained description of the model is presented here, making explicit the underlying assumptions. 


\subsection{Active galactic nuclei accretion disc}

We consider a geometrically thin and optically thick accretion disc following Keplerian rotation around a central SMBH with mass $M_\textrm{SMBH}$, meaning the orbital frequency of the disc is  $\Omega(r) = \sqrt{\frac{GM_\mathrm{SMBH}}{r^3}}$. Angular momentum is transferred as a result of friction between adjacent layers, and the resulting viscous torque acting on the gas is given by \citep{Pringle_1981} 
\begin{equation}
T_{visc}(r) = 2 \pi r \nu \Sigma_g r^2 \frac{d\Omega}{dr} , 
\end{equation}
where $\nu$ is the kinematic viscosity and $\Sigma_g$ is the gas surface density. 
As $\Omega(r)$ decreases outwards for Keplerian rotation, this torque results in an outward transport of angular momentum. 

The actual nature and magnitude of the accretion disc viscosity is still unclear, but it can be parametrised by the well-known $\alpha$-prescription \citep{Shakura_Sunyaev_1973} 
\begin{equation}
\nu = \alpha c_s H , 
\end{equation}
where $0 < \alpha < 1$ is a dimensionless parameter, $c_s$ is the local sound speed, and $H$ is the disc scale height.  
Assuming hydrostatic equilibrium in the vertical direction, $H/r \sim c_s/v_{\phi}$ (where $v_{\phi} = r \Omega$ is the circular velocity), the sound speed can also be expressed as
\begin{equation}
c_s = H \Omega 
= h \sqrt{\frac{G M_\mathrm{SMBH}}{r}} , 
\end{equation}
where $h = H/r$ is the disc aspect ratio. 
For a geometrically thin disc we have $h = H/r \ll 1$.
Combined with the above-introduced $\alpha$-prescription, the viscous torque can be written as
\begin{equation}
T_{visc}(r) = - 3 \pi \alpha c_s^2(r) \Sigma_g(r) r^2 .
\label{Eq_Tvisc}
\end{equation} 

Assuming an isothermal distribution, the total mass enclosed within a radius $r$ is given by $M(r) = \frac{2 \sigma^2 r}{G}$, where $\sigma$ is the velocity dispersion. We further assume that the gas mass is a fraction $f_g$ of the total mass, that is, $M_g(r) = f_g M(r)$. The gas surface density is then given by $\Sigma_g = \frac{f_g \sigma^2}{\pi G r}$. 
The stellar velocity dispersion can be related to the SMBH mass via the empirical $M - \sigma$ relation \citep{Kormendy_Ho_2013} 
\begin{equation}
\frac{M_\text{SMBH}}{10^9 M_{\odot}} = (0.31^{+0.037}_{-0.033}) \left( \frac{\sigma}{200 \, \mathrm{km s^{-1}} }\right)^{4.38\pm0.29} . 
\end{equation}


\subsection{Disc-binary interaction}
\label{Subsec_DB_interaction}

We consider a BBH system embedded in the AGN gaseous disc. 
The binary is of total mass $M_b = m_1 + m_2$ on an elliptic orbit with semi-major axis $a$ and eccentricity $e$, and we work throughout in the reference frame of the binary system. For the binary system not to be disrupted by gravitational torques of the SMBH, we require that the binary does not cross the Roche limit of the central mass. This stability criterion is given by  $\frac{r}{a} \gtrsim (1+e) ( \frac{3 M_\textrm{SMBH}}{M_b} )^{1/3}$, where $r$ is the radial distance of the binary from the central mass  \citep{Hoang_et_2018,Deme2020}. 
Gravitational torques from the binary may clear a cavity in the surrounding gas distribution, as suggested by numerical simulations \citep{Artymowicz_Lubow_1994, Armitage_Natarajan_2005}. We assume that the binary resides at the centre of the inner depleted cavity, which is surrounded by a circumbinary disc (CBD). 
A rough estimate for the gap to open is the condition that tidal torques from the binary overwhelm viscous torques in the disc. This means  that the disc scale height $H$ of the CBD be smaller than the Hill radius $R_H = r \left( \frac{M_b}{3 M_\mathrm{SMBH}} \right)^{1/3}$ of the binary system. 
Accordingly, the disc aspect ratio should be smaller than $h < \left( \frac{M_b}{3 M_\mathrm{SMBH}} \right)^{1/3}$.
In principle, gas flows from the CBD can penetrate within the cavity, leading to the formation of mini-discs around the individual black holes. 
For simplicity, here we assume that the accretion rates are suppressed, so the cavity is depleted and neither the primary nor the secondary BH have their own mini-discs. 
 We will discuss how the inclusion of accretion flows within the cavity may affect the binary orbital evolution in Sect. $\ref{Subsect_accretion}$. 
 
The total energy of the BBH is given by
\begin{equation}
E_b = - \frac{G M_b \mu}{2a} ,
\label{Eq_Eb}
\end{equation}
where $\mu = \frac{m_1 m_2}{M_b} = \frac{q}{(1+q)^2} M_b$ is the reduced mass, and $q = m_2/m_1$ is the mass ratio. 
The orbital angular momentum of the BBH is given by
\begin{equation}
L_b = \mu a^2 \Omega_b \sqrt{1-e^2} ,
\label{Eq_Lb}
\end{equation}
where $\Omega_b = \sqrt{\frac{G M_b}{a^3}}$ is the orbital frequency of the binary. Differentiating Eq. \ref{Eq_Eb} with respect to time, and assuming that the accretion rates onto the individual BHs are negligible ($\dot{m_1} = \dot{m_2} = 0$), one obtains 
\begin{equation}
\frac{\dot{E}_b}{E_b} = - \frac{\dot{a}}{a} . 
\end{equation}
Similarly, the rate of change of the orbital angular momentum is given by
\begin{equation}
\frac{\dot{L}_b}{L_b} = \frac{\dot{a}}{2a} - \frac{e \dot{e}}{(1-e^2)} ;
\end{equation}
here we have again neglected relative and total accretion rates ($ \dot{\mu} = \dot{M} = 0$). 

The evolution of the semi-major axis $a$ and of the orbital eccentricity $e$ of the binary are driven by tidal and viscous interactions between the binary and the CBD. 
The microscopic dynamics of the energy and angular momentum exchange between the binary and its CBD are complex, and we employ idealised assumptions. We first of all assume that the disc-binary interaction may well be approximated as an adiabatic process. This means that the characteristic timescale of the disc-binary interaction is much longer than some characteristic timescales of the binary, for example its orbital period. We furthermore assume that the non-axisymmetric potential perturbations of the binary system are small around the average binary potential. Under these circumstances, the binary energy dissipation rate is related to its change in angular momentum through the orbital frequency (see e.g. \cite{LeTiec2015})
\begin{equation}
\dot{E}_b = \Omega_b \dot{L}_b . 
\end{equation}
 
The assumption that torques on average act axisymmetrically upon the disc goes hand in hand with the assumption that the shape of the cavity is circular and remains circular throughout the orbital evolution. In more realistic situations, the CBD can be distorted and become eccentric as a result of its interaction with the binary. 
Indeed, the development of accretion streams within the cavity may affect the shape of the disc and drive the disc eccentricity growth. Based on two-dimensional (2D) hydrodynamic simulations, \citet{MacFadyen_Milosavljevic_2008} show that an initially circular disc can become eccentric. Only a small fraction of the accretion streams generated at the inner edge of the disc actually reach the individual BHs. The remaining gas streams are flung back towards the disc, where they contribute to the deformation of the disc shape \citep{DOrazio_et_2013}. The growth of the disc eccentricity through such stream impact has also been observed in three-dimensional (3D) magnetohydrodynamic (MHD) simulations of an equal-mass binary on a circular orbit \citep{Shi_et_2012}. 
Therefore numerical simulations show that the disc eccentricity can be induced even for circular binaries \citep[see also][]{Farris_et_2014}. 
In the case of eccentric binaries, the disc eccentricity growth can be due to direct driving \citep{Lubow_Artymowicz_2000}. 
Two-dimensional hydrodynamical simulations suggest that cavities tend to become eccentric in most cases \citep{Moesta_et_2019}. We do not consider eccentric discs in our analytic modelling and assume a circular cavity shape \citep[see also][]{Hayasaki_2009}. In the following, we denote the inner edge of the circular cavity by $r_{in}$, although no sharp edge is expected due to stream impact and the presence of shocked material. 

From Eqs. \ref{Eq_Eb} and \ref{Eq_Lb} one observes that the binary orbital energy and angular momentum are related by
\begin{equation}
E_b = - \frac{\Omega_b L_b}{2 \sqrt{1-e^2}}. 
\end{equation}

As a result, the semi-major axis evolution can be written as
\begin{equation}
\frac{\dot{a}}{a} = 2 \frac{\dot{L}_b}{L_b} \sqrt{1-e^2} .
\label{Eq_adot_Lb}
\end{equation} 
Similarly, the expression for the orbital eccentricity evolution can be written
\begin{equation}
\frac{\dot{e} e}{1-e^2} = \frac{\dot{L}_b}{L_b} (\sqrt{1-e^2} - 1) .
\label{Eq_edot_Lb}
\end{equation} 

Conservation of angular momentum implies that the orbital angular momentum deposited by the binary is completely absorbed by the surrounding gas disc. The angular momentum is transferred through the inner edge $r_{in}$ of the CBD, that is, $- \dot{L}_b = T_{visc} (r_{in})$. In the binary reference frame, we have $\dot{L}_b < 0$ and accordingly the sign of the viscous torque must be opposite to that of Eq. \ref{Eq_Tvisc}. 

Numerical simulations suggest that the size of the central cavity typically extends to about twice the semi-major axis for binaries on near-circular orbits. 
The location of the inner edge depends on the orbital eccentricity, and $r_\text{in}$ progresses outwards with increasing eccentricity: for instance, smoothed particle hydrodynamic (SPH) simulations indicate that the inner edge shifts from $r_{\text{in}} \approx 1.9 a$ for $e \approx 0.02$ to $r_{\text{in}} \approx 3 a$ for $e \approx 0.6$ \citep{Artymowicz_Lubow_1994}. 
Recent 2D hydrodynamical simulations also show that the radial extent of the cavity increases with increasing eccentricity \citep{Moesta_et_2019}.  
Inspired by these simulations, we estimate the inner edge of the CBD to be located at twice the distance of the binary at apocentre, that is, $r_{\text{in}} = 2 a (1+e)$. As a consequence, the viscous torque (see Eq. \ref{Eq_Tvisc}) at the inner edge of the CBD is given by $T_{visc}(r_{in}) = 12 \pi \alpha c_s^2 \Sigma_g a^2 (1+e)^2$. 

Inserting the corresponding values of $\dot{L}_b$ and $L_b$ into Eqs. \ref{Eq_adot_Lb} and \ref{Eq_edot_Lb}, we obtain the evolution equations in the disc-driven regime: 
\begin{equation}
\frac{\dot{a}}{a} =  -\frac{24 \pi \alpha c_s^2 \Sigma_g (1+e)^2}{\mu \Omega_b} , 
\label{Eq_a_disc}
\end{equation}
\begin{equation}
\frac{\dot{e} e}{1-e^2} = \frac{12 \pi \alpha c_s^2 \Sigma_g (1+e)^2 }{\mu \Omega_b} \left(\frac{1}{\sqrt{1-e^2}} -1 \right) .
\label{Eq_e_disc}
\end{equation} 

The disc-driven evolution Eqs. \ref{Eq_a_disc} and \ref{Eq_e_disc} can be combined to give
\begin{equation}
\frac{\dot{e} e}{1-e^2} = \frac{1}{2} \frac{\dot{a}}{a} \left(1- \frac{1}{\sqrt{1-e^2}} \right) . 
\label{Eq_e_a_comb}
\end{equation}

By dividing Eq. \ref{Eq_a_disc} by Eq. \ref{Eq_e_disc}, and solving the resulting differential equation we get
\begin{equation}
a(e) = a_0 \left( \frac{1 - \sqrt{1-e_0^2}}{1 - \sqrt{1-e^2}} \right)^2 , 
\label{Eq_a_e}
\end{equation}
where $a_0$ and $e_0$ are the initial semi-major axis and initial eccentricity, respectively. 
Starting from given initial conditions ($a_0, e_0$), the disc-binary dynamics outlined here will drive the binary into a regime where gravitational wave emission will dominate the orbital evolution (see Sect. \ref{Subsection_GWinspiral}). 


\subsection{Gravitational-wave-driven inspiral}
\label{Subsection_GWinspiral}

When the orbital separation of a BBH system is sufficiently small, the effects of GW emission on the orbital evolution cannot be neglected. We recall the GW-driven evolution equations of the orbital elements \citep{Peters_1964} 
\begin{equation}
\dot{a} = - \frac{64}{5} \frac{G^3}{c^5} \frac{\mu M_b^2}{a^3} \frac{1}{(1-e^2)^{7/2}} \left( 1 + \frac{73}{24} e^2 + \frac{37}{96} e^4 \right),
\label{Eq_a_GW}
\end{equation}
\begin{equation}
\dot{e} = -  \frac{304}{15} \frac{G^3}{c^5} \frac{\mu M_b^2}{a^4} \frac{e}{(1-e^2)^{5/2}} \left( 1 + \frac{121}{304} e^2 \right). 
\label{Eq_e_GW}
\end{equation} 

Because of the eccentricity enhancement factor, the GW inspiral can be considerably accelerated for high eccentricities. As an example, $\dot{a}_\text{GW}$ may increase by a factor of $\sim 10^3$ from $e = 0$ to $e = 0.9$.


\section{Coupled disc+GW-driven evolution}
\label{Sect_coupled_evol}

Both gravitational wave emission and disc-binary interaction affect the overall evolution of the BBH system in the AGN accretion disc. In order to investigate the interplay between the two effects, we combine the disc-driven (Eqs. \ref{Eq_a_disc}-\ref{Eq_e_disc}) with the corresponding GW-driven (Eqs. \ref{Eq_a_GW}-\ref{Eq_e_GW}) evolution equations. The resulting coupled disc+GW-driven evolution is described by
\begin{equation}
\dot{a} =  \dot{a}_\text{disc} + \dot{a}_\text{GW},
\label{Eq_a_coupled}
\end{equation}
\begin{equation}
\dot{e} =  \dot{e}_\text{disc} + \dot{e}_\text{GW}.
\label{Eq_e_coupled}
\end{equation}

Numerical integration of this coupled system of differential equations yields the temporal evolution of the semi-major axis $a(t)$ and orbital eccentricity $e(t)$. We note that the values of the CBD parameters ($\alpha$, $c_s$, $\Sigma_g$) should be in principle estimated at the inner edge $r_{in}$. Here we assume that these values are not greatly different and reflect the local values of the background AGN disc at the radial location of the binary (as in \citet{Baruteau_et_2011}). 
In order to gain a better insight, we develop the respective expressions of the evolution equations and present below the explicit dependence on the underlying parameters. The resulting analytic scalings in the disc-driven regime are given by
\begin{equation}
\dot{a}_\text{disc} \propto - \alpha f_g h^2 \frac{M_\text{SMBH}^{3/2}}{r^2} \frac{(1+q)^2}{q} M_b^{-3/2} a^{5/2} (1+e)^2 , 
\label{Eq_disc_analy_a}
\end{equation}
\begin{equation}
\dot{e}_\text{disc} \propto \alpha f_g h^2 \frac{M_\text{SMBH}^{3/2}}{r^2} \frac{(1+q)^2}{q} M_b^{-3/2} a^{3/2} \frac{(1+e)^2}{e} (\sqrt{1-e^2} +e^2 -1) . 
\label{Eq_disc_analy_e}
\end{equation} 
In an analogous way, the scalings for the GW-driven regime are given by
\begin{equation}
\dot{a}_\text{GW} \propto - \frac{q}{(1+q)^2} M_b^3 a^{-3} \frac{1}{(1-e^2)^{7/2}} \left( 1 + \frac{73}{24} e^2 + \frac{37}{96} e^4 \right) , 
\label{Eq_GW_analy_a}
\end{equation}
\begin{equation}
\dot{e}_\text{GW} \propto - \frac{q}{(1+q)^2} M_b^3 a^{-4} \frac{e}{(1-e^2)^{5/2}} \left( 1 + \frac{121}{304} e^2 \right) . 
\label{Eq_GW_analy_e}
\end{equation}  

These analytic expressions will be useful when interpreting the dependence of the merger timescale and eccentricity evolution on the underlying physical parameters (Sect. \ref{Sect_mergertime}-\ref{Sect_ecc_evol}). 


\section{Binary orbital evolution in active galactic nuclei discs}
\label{Sect_bin_evol}

We now analyse the evolution of the semi-major axis $a$ and the orbital eccentricity $e$ that is dictated by the coupled disc+GW-driven evolution Eqs. (\ref{Eq_a_coupled}) and (\ref{Eq_e_coupled}).

\subsection{Semi-major axis decay}

\begin{figure}[h!]
  \centering
 \includegraphics[width=0.45\textwidth]{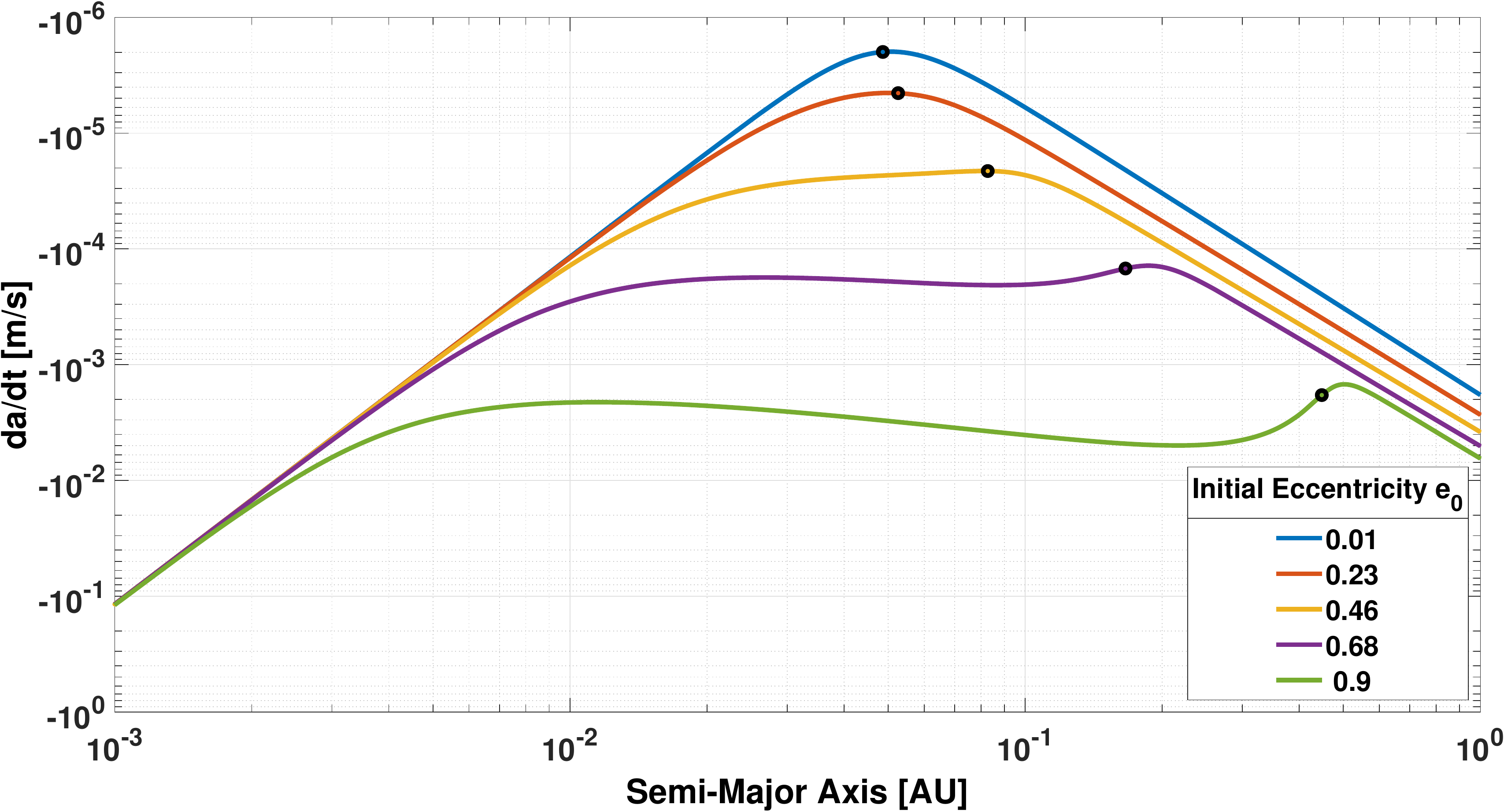}
    \caption{Orbital decay rate $\dot{a}$  displayed as a function of the orbital separation $a$ for different initial eccentricities ($e_0 = 0.01 - 0.9$). The orbital and disc parameters are set to the fiducial values stated in Sect. 4.1. The black circles indicate the transition from the disc-driven to the GW-driven regime of the evolution channel. }
    \label{davsa_e0}
\end{figure}

Figure \ref{davsa_e0} shows the rate of orbital decay as a function of the orbital separation for different initial eccentricities. 
The following values are taken as fiducial parameters of the accretion disc model: viscosity parameter $\alpha = 0.1$, gas fraction $f_g =0.1$, and aspect ratio $h = H/r = 0.01$. The binary parameters are set to: $a_0 = 1$ AU, $M_b = 50 M_\odot$, and $q=1$. 
In this fiducial model, the binary is located at a distance of $r = 0.1$ pc in an AGN disc surrounding a SMBH of mass $M_{\text{SMBH}} = 10^7 M_\odot$. 
Unless not subject to variation, we will use these fiducial parameter values throughout the text for plots and numerical results. We note that the stability criterion of Sect. \ref{Subsec_DB_interaction} is fulfilled for the fiducial model parameters. Furthermore, the condition for gap opening of Sect. \ref{Subsec_DB_interaction} entails a thin disc with $h \lesssim 0.01$, which is in accordance with our fiducial aspect ratio. The purpose of aspect ratio sweeps (like in Table \ref{Tab_Table}), which go up to $h = 0.1,$ is to illustrate relative dependences. We further discuss the important role of the disc aspect ratio on the disc-binary interaction in Sect. \ref{Subsect_accretion}. 
 
In Fig. \ref{davsa_e0} we observe two distinct trends corresponding to the disc-driven regime at large separations and the GW-driven regime at small separations, respectively. 
The black circles indicate the semi-major axis where the binary transitions from the disc-dominated regime into the GW-dominated regime, that is, where $\dot{a}_\text{disc} = \dot{a}_\text{GW}$. 
The binary semi-major axis continuously decreases: the rate of orbital decay is driven by the disc-binary interaction at large $a$, while it is dominated by GW emission at small $a$ (reaching a minimal value around the transition point). 
We note that the rate of orbital decay in the disc-driven regime $\dot{a}_\text{disc}$ decreases with decreasing separation, but the GW-driven orbital decay rate $\dot{a}_\text{GW}$ increases with decreasing separation instead.
These physical trends can be directly seen from the analytic scalings. 
In the purely disc-driven regime we find from solving Eqs. \ref{Eq_a_disc} and \ref{Eq_a_e} that the semi-major axis evolves as
\begin{equation}
\dot{a}_{\text{disc}} \propto  - a^{5/2} \left( 1 + a^{- 1/4} \right)^2 .
\end{equation}
On the other hand, in the purely GW-driven regime, the rate of orbital decay simply scales as
 \begin{equation}
\dot{a}_{\text{GW}} \propto  - a^{-3},
\end{equation}
since the orbit has nearly circularised for a semi-major axis below $a \lesssim 10^{-3}$AU. 
From Fig. \ref{davsa_e0}, we also note that the critical transition points occur at a larger semi-major axis for higher initial eccentricities. This implies that for a higher $e_0$, the binary transitions earlier into the GW-regime (that is, at a larger separation) where the high eccentricity eventually leads to a faster inspiral (due to the steep dependence on eccentricity, cf. Eqs. \ref{Eq_a_GW}-\ref{Eq_e_GW}).

\subsection{ Orbital eccentricity growth}

\begin{figure}[h!]
  \centering
 \includegraphics[width=0.45\textwidth]{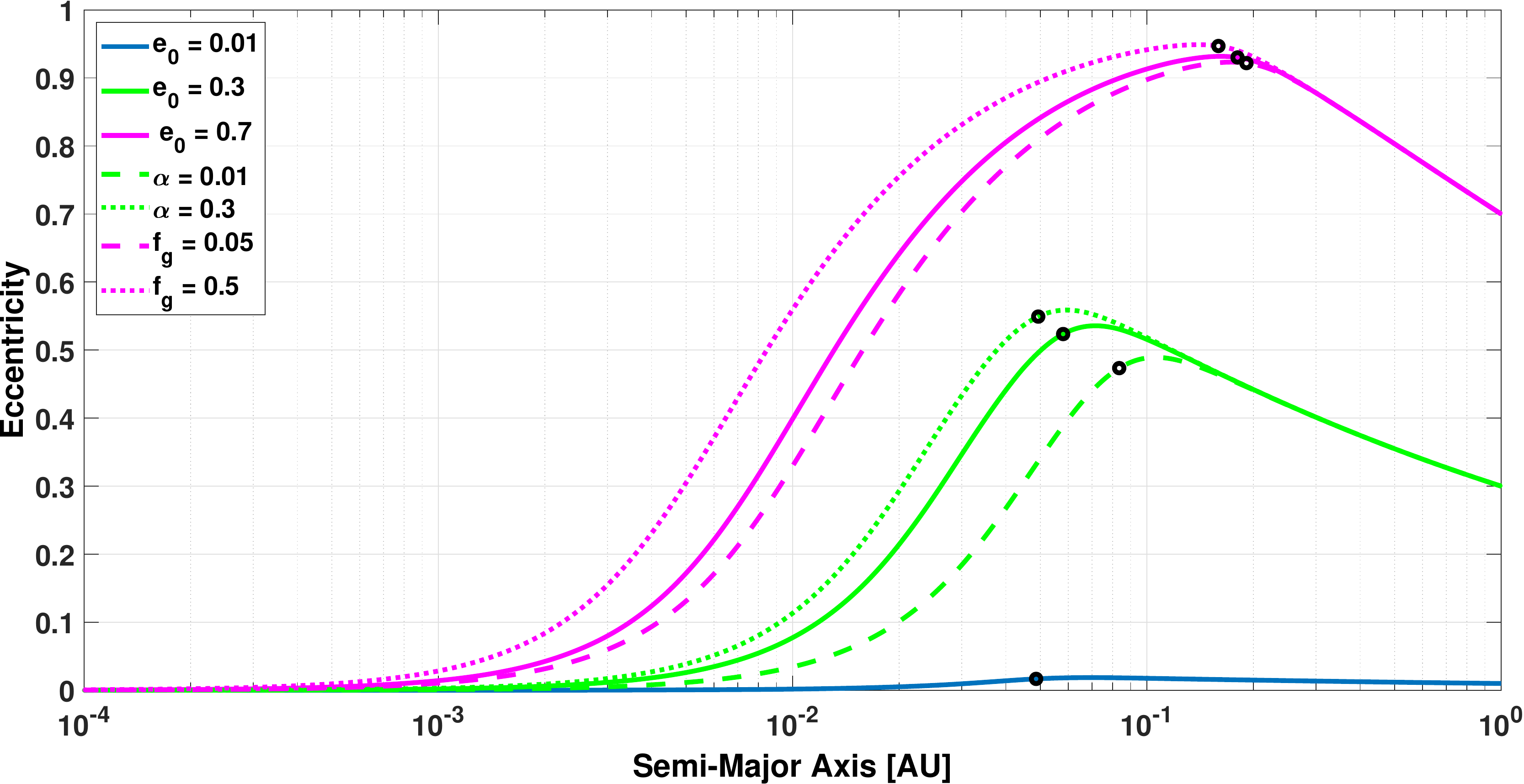}
    \caption{Eccentricity plotted as a function of the semi-major axis for various disc and binary configurations (cf. Sect. 4.1 for the fiducial parameter values). The blue line uses $e_0 = 0.01$ for the initial eccentricity, the green  $e_0 = 0.3$, and the magenta $e_0 = 0.7$. The viscosity parameter $\alpha$ is set to $\alpha = 0.01$ in the green dashed line and to $\alpha = 0.3$ in the green dotted line. The gas fraction $f_g$ is set to $f_g = 0.05$ in the magenta dashed line and to $f_g = 0.5$ in the magenta dotted line.}
    \label{evsa_e0alfg}
\end{figure}

In Fig. \ref{evsa_e0alfg} we plot the orbital eccentricity as a function of the semi-major axis for variations in the initial eccentricity $e_0$, viscosity parameter $\alpha$, and gas fraction $f_g$. 
We note that $f_g \sim 0.16$ is the cosmological gas fraction, and a standard value of $f_g \sim 0.1$ is adopted in many studies (e.g. \cite{King_Pounds_2015}). For instance, \citet{Thompson_et_2005} assume a fiducial gas fraction of $f_g \sim 0.1$ for their accretion disc model. Higher values ($f_g \sim 0.5$) may be expected in gas-rich galaxies at higher redshifts \citep[e.g.][]{Daddi_et_2010}, while lower values are more likely in gas-depleted galaxies. 
 Observational estimates of the disc viscosity parameter suggest values in the range $\alpha \sim (0.1 - 0.4)$, while smaller values ($\alpha < 0.01$) seem to be favoured by numerical simulations  \citep{King_et_2007, Martin_et_2019}. 
Here we consider viscosity parameters and gas fractions within plausible ranges of $0.01 \leq \alpha \leq 0.3$ and $0.05 \leq f_g \leq 0.5$. 
We observe that the orbital eccentricity grows in the disc-driven regime and decays in the GW-driven regime, while the semi-major axis decreases in both regimes. Thus the disc-binary interaction shrinks the orbital separation while at the same time it increases the orbital eccentricity. In the case of a small initial eccentricity ($e_0 = 0.01$, blue curve), the subsequent eccentricity growth in the disc-driven regime is almost negligible. In contrast, the eccentricity growth is more prominent for larger initial eccentricities, and with increasing $e_0$ the binary transitions into the GW regime with a  higher eccentricity. From Fig. \ref{evsa_e0alfg} we also note that increasing the viscosity parameter $\alpha$ and the gas fraction $f_g$ leads to larger eccentricity growth, with the binary reaching higher maximal values $e_\text{max}$. For a given initial eccentricity, the critical transition points shift to smaller separations for larger viscosity parameters and gas fractions. 

\begin{figure}[h!]
  \centering
   \includegraphics[width=0.45\textwidth]{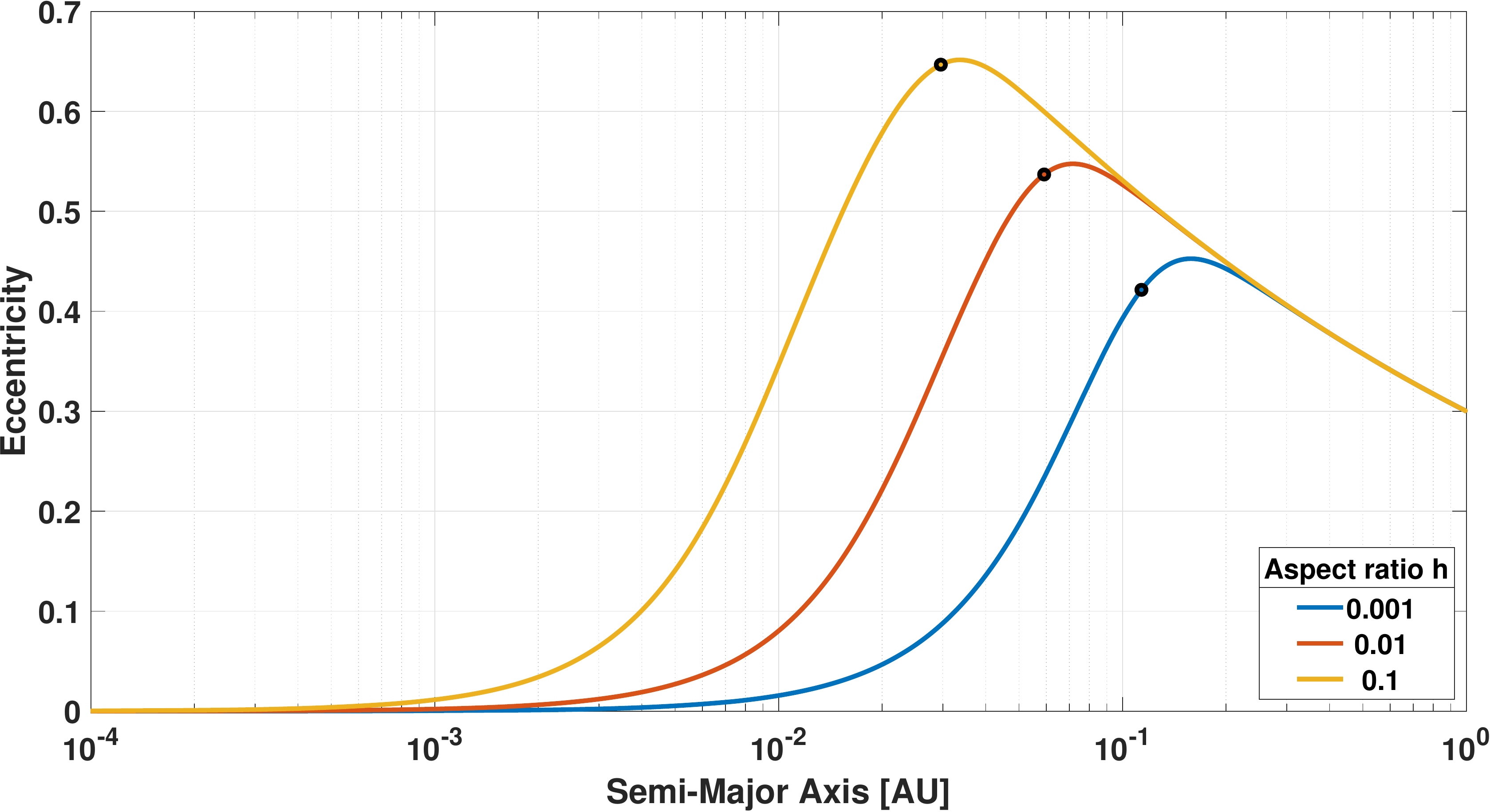}
    \caption{Orbital eccentricity shown as a function of the semi-major axis for different aspect ratios ($h = 0.001, 0.01, 0.1$).}
    \label{evsa_h}
\end{figure}

In Fig. \ref{evsa_h} we show an analogous plot of the orbital eccentricity versus the semi-major axis, for different values of the disc aspect ratio $h = H/r$. As previously noted, the eccentricity grows in the disc-driven regime (at large separations), while it decays in the GW-driven regime (at small separations). We see that for larger aspect ratios, the eccentricity growth can be significantly larger, reaching higher  $e_\text{max}$ values. The transition points also shift to smaller separations for larger $h$. 
We note that the slope in the disc-driven regime is roughly constant for different aspect ratios. 
In fact, in the disc-driven regime we obtain from Eq. \ref{Eq_e_a_comb}
\begin{equation}
a \frac{de}{da} = \frac{1}{2} \frac{1-e^2}{e} \left( 1 - \frac{1}{\sqrt{1-e^2}} \right) , 
\end{equation}
which is only a function of the eccentricity and does not depend explicitly on the accretion disc parameters.  

The physical trends observed in Figs. \ref{evsa_e0alfg}-\ref{evsa_h} and the dependence on the underlying accretion disc parameters can be interpreted in terms of simple analytic scalings. Indeed, both the rate of orbital decay and eccentricity growth in the disc-driven regime scale as $\propto \alpha f_g h^2$, with a quadratic dependence on the disc aspect ratio. 
N-body SPH simulations also suggest that an increase in the disc aspect ratio leads to both enhanced orbital decay and eccentricity growth \citep{Fleming_Quinn_2017}.

\section{Quantifying the binary black hole merger timescale}
\label{Sect_mergertime}

We next quantify the merger timescale of BBHs embedded in AGN accretion discs. Based on the coupled disc+GW-driven evolution equations (Eqs. \ref{Eq_a_coupled}-\ref{Eq_e_coupled}), we derive the merger time $\tau_\text{merger}$, defined as the point where the numerical solution $a(t)$, or equivalently $e(t)$, approaches the abscissa axis. 
In general, the merger timescale is a function of both the accretion disc and binary orbital parameters: $\tau_\text{merger}(\alpha, f_g, h, r, M_\text{SMBH}, a_0, e_0, q, M_b)$. Below we analyse its dependence on the different quantities. 

\begin{figure}[h!]
  \centering
 \includegraphics[width=0.45\textwidth]{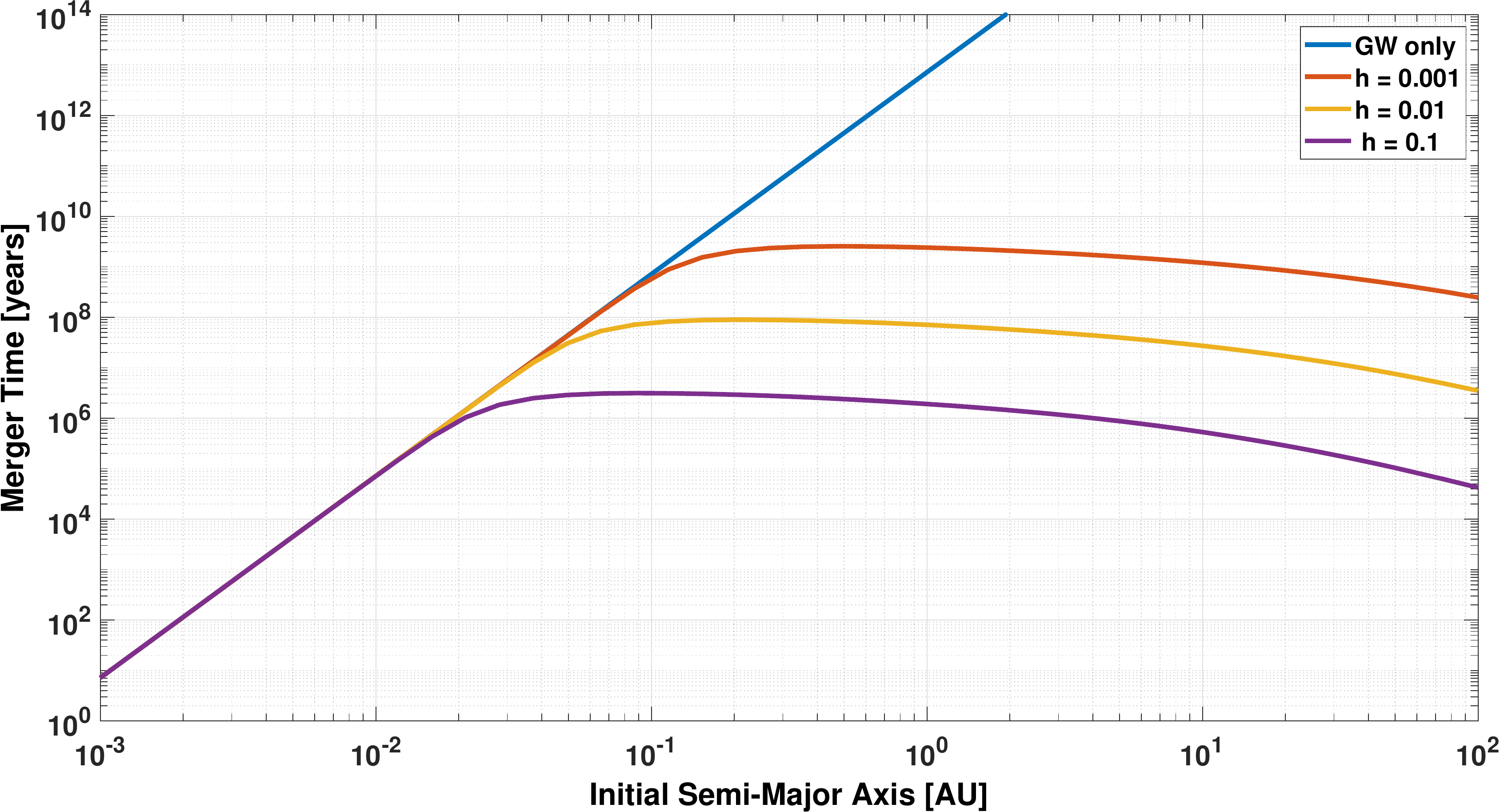}
    \caption{Merger timescale $\tau_\text{merger}$ as a function of initial semi-major axis $a_0$. The blue line is the merger time expected from purely GW-emission; it does not take into account the disc-binary interaction. The other curves include the disc-binary interaction. The disc aspect ratio is set to $h=0.001$ (red), $h=0.01$ (yellow), and $h = 0.1$ (purple). }
    \label{mrgtime}
\end{figure}

Figure $\ref{mrgtime}$ shows the merger timescale as a function of the initial orbital separation for GW-only evolution, and including the disc-driven evolution with varying disc aspect ratios. 
We recall that the orbital decay is more efficient at large separations in the disc-driven regime, while it is more efficient at small separations in the GW regime (suggesting that the bottleneck is located around the transition point). 
In the case of purely GW-driven decay, the merger time exceeds the Hubble time ($t_\text{H} \sim 10^{10}$ yr), unless the initial separation is smaller than $a_0 \lesssim 2 \times 10^{-1}$ AU. In fact, the merger time is already $\sim 7 \times 10^{12}$ yr at $a_0 = 1$ AU, indicating that binaries cannot merge within the age of the Universe. 
We also observe that the merger timescale in the GW-driven regime steeply increases with increasing separation as $\tau_\text{GW} \sim \frac{a}{\dot{a}_\text{GW}} \propto a^4$. In contrast, when including the disc-binary interaction, the merger time can be considerably reduced. It falls in the range $\tau_\text{merger} \sim (10^6 - 10^9)$ yr, hence allowing BBH mergers to occur on astrophysically plausible timescales. For our fiducial model with a disc aspect ratio of $h = 0.01$, the corresponding merger time is $\tau_\text{merger} \sim 7 \times 10^7$ yr at $a_0 = 1$ AU, so the timescale is reduced by five orders of magnitude compared to the GW-only decay. Furthermore, within the disc-dominated regime, the decay timescale decreases with increasing separation, roughly following a scaling of the form $\tau_\text{disc} \sim \frac{a}{\dot{a}_\text{disc}} \propto a^{-3/2}$ (to leading order). 

From Fig. \ref{mrgtime} we see that an increase in the disc aspect ratio can lead to a considerable decrease in the merger timescale. 
For instance, at an initial separation of 1 AU, the merger time is $\tau_\text{merger} \sim 2 \times 10^9$ yr for $h = 10^{-3}$, while it is $\tau_\text{merger} \sim 2 \times 10^6$ yr for $h = 0.1$, thus varying by nearly three orders of magnitude. This is likely due to the quadratic dependence of the orbital decay on the disc aspect ratio (Eqs. \ref{Eq_disc_analy_a}-\ref{Eq_disc_analy_e}).  
In an analogous way, the merger timescale can also vary significantly depending on the other underlying parameters. 
In Table \ref{Tab_Table} we summarise the dependence of the merger time on the different accretion disc and binary orbital parameters, quoting alongside the maximal eccentricity reached ($e_\text{max}$) and the corresponding semi-major axis ($a(e_\text{max})$). 
From Table \ref{Tab_Table}, we see that an increase in the viscosity parameter and gas fraction leads to a reduction in the merger timescale. Likewise, an increase in the SMBH mass and a decrease in the radius also lead to shorter merger times. 
Thus a favourable combination of the AGN accretion disc parameters (large $\alpha, f_g, h, M_\text{SMBH}$ and small $r$) can yield the shortest merger times, and as a consequence rapid BBH mergers. 
With regard to the binary orbital parameters, we see that a higher initial eccentricity, larger initial semi-major axis, smaller total binary mass, and smaller mass ratio lead to shorter merger timescales.
Concerning the latter scaling, we note that the binary orbital evolution ($\dot{a}$, $\dot{e}$) has an opposite dependence on the mass ratio ($q$) in the disc-driven and GW-driven regimes (see Eqs. $\ref{Eq_disc_analy_a} - \ref{Eq_GW_analy_e}$). 
The orbital decay is faster for smaller mass ratios in the disc-driven regime while it is slower in the GW-driven regime. The former trend dominates the global disc+GW evolution, leading to overall shorter merger timescales.  
Obviously, variations in the AGN disc parameters only affect the evolution in the disc-driven regime; whereas the binary orbital parameters appear in both disc-driven and GW-driven regimes (with sometimes opposite dependences), making the interpretation a little more subtle (see also Sect. \ref{Sect_ecc_evol}). 

\begin{center}
\begin{table}
\begin{tabular}{cc|c|c|c|c|l}
\cline{3-5}
& & $e_{\text{max}}$ & $a(e_{\text{max}})$ [AU] & $\tau_\text{merger}$ [yr]   \\ \hline \cline{1-5}
\multicolumn{1}{ |c  }{\multirow{2}{*}{$\alpha$} } &
\multicolumn{1}{ |c| }{$ 0.01$} & $ 0.49$ & $0.11$ & $4.2 \times 10^8$    \\ \cline{2-5}
\multicolumn{1}{ |c  }{}                        &
\multicolumn{1}{ |c| }{$0.3$ } & $ 0.56$ & $0.06$ & $3.0 \times 10^7$    \\ \hline \cline{1-5}
\multicolumn{1}{ |c  }{\multirow{2}{*}{$f_g$} } &
\multicolumn{1}{ |c| }{$0.05$} & $0.52$ & $0.08$ & $1.2 \times 10^8$    \\ \cline{2-5}
\multicolumn{1}{ |c  }{}                        &
\multicolumn{1}{ |c| }{$0.5$} & $0.57$ & $ 0.05$ & $ 2.0 \times 10^7$    \\ \hline  \cline{1-5} 
\multicolumn{1}{ |c  }{\multirow{2}{*}{$h$} } &
\multicolumn{1}{ |c| }{$0.001$} & $ 0.45$ & $0.16$ & $2.4 \times 10^9$    \\ \cline{2-5}
\multicolumn{1}{ |c  }{}                        &
\multicolumn{1}{ |c| }{$0.1$} & $ 0.64$ & $0.03$ & $ 1.9 \times 10^6$    \\ \hline  \cline{1-5} 
\multicolumn{1}{ |c  }{\multirow{2}{*}{r} } &
\multicolumn{1}{ |c| }{$0.01$ pc} & $0.74$ & $0.02$ & $3.9 \times 10^4$    \\ \cline{2-5}
\multicolumn{1}{ |c  }{}                        &
\multicolumn{1}{ |c| }{$1$ pc} & $0.54$ & $ 0.07$ & $6.0 \times 10^7$    \\ \hline  \cline{1-5} 
\multicolumn{1}{ |c  }{\multirow{2}{*}{$M_\text{SMBH}$} } &
\multicolumn{1}{ |c| }{$ 10^6 \, M_{\odot}$} & $0.47$ & $0.13$ & $9.3 \times 10^8$    \\ \cline{2-5}
\multicolumn{1}{ |c  }{}                        &
\multicolumn{1}{ |c| }{$ 10^9 \, M_{\odot}$} & $0.68$ & $ 0.02$ & $3.6 \times 10^5$    \\ \hline  \cline{1-5}  
\multicolumn{1}{ |c  }{\multirow{2}{*}{$e_0$} } &
\multicolumn{1}{ |c| }{$0.01$} & 0.02 & 0.07 & $1.8 \times 10^8$   \\ \cline{2-5}
\multicolumn{1}{ |c  }{}                        &
\multicolumn{1}{ |c| }{$0.9$} & 0.98 & 0.36 & $1.6 \times 10^6$    \\   \hline  \cline{1-5} 
\multicolumn{1}{ |c  }{\multirow{2}{*}{$a_0$} } &
\multicolumn{1}{ |c| }{$1$ AU} & 0.54 & 0.07 & $7.1 \times 10^7$    \\ \cline{2-5}
\multicolumn{1}{ |c  }{}                        &
\multicolumn{1}{ |c| }{$20$ AU} & 0.88 & 0.12 & $1.7 \times 10^7$  \\ \hline  \cline{1-5} 
\multicolumn{1}{ |c  }{\multirow{2}{*}{$q$} } &
\multicolumn{1}{ |c| }{$0.05$} & $0.61$ & $0.04$ & $2.7 \times 10^7$    \\ \cline{2-5}
\multicolumn{1}{ |c  }{}                        &
\multicolumn{1}{ |c| }{$1$} & $0.54$ & $ 0.07$ & $7.1 \times 10^7$    \\ \hline  \cline{1-5} 
\multicolumn{1}{ |c  }{\multirow{2}{*}{$M_b$} } &
\multicolumn{1}{ |c| }{$10 \, M_{\odot}$} & 0.69 & 0.02 & $2.9 \times 10^7$   \\ \cline{2-5}
\multicolumn{1}{ |c  }{}                        &
\multicolumn{1}{ |c| }{$100 \, M_{\odot}$} & 0.47 &  0.12 & $ 9.7 \times 10^7$   \\ \hline \cline{1-5} \\
\end{tabular} 
  \caption{Maximal eccentricity $e_{\text{max}}$, the corresponding semi-major axis $a(e_{\text{max}})$, and the time to coalescence $\tau_\text{merger}$ for parameter configurations where only a selected AGN or binary parameter is varied (for the unvaried parameters the fiducial values of Sect. 4.1 are adopted).}
   \label{parameterstudy} 
  \label{Tab_Table}
\end{table}
\end{center}


\section{Eccentricity evolution in the LISA band } 
\label{Sect_ecc_evol}

In this section we explore the implications of the peculiar orbital eccentricity evolution resulting from the coupled disc+GW evolution on the GW frequency signal. 
We take the peak rest-frame GW frequency as
\begin{equation}
f_\text{GW} \approx 2 \Omega_b (1 - e)^{-3/2};
\label{Eq_e_f}
\end{equation} 
at this frequency most of the GW power is emitted \citep[e.g.][]{Dorazio_Samsing_2018}. 
Due to the eccentricity dependence of the above equation, for $e > 0$ the GW frequency can be significantly higher than the orbital frequency, shifting the binaries towards a frequency range more easily detectable by LISA.

\begin{figure}[h!]
  \centering
 \includegraphics[width=0.45\textwidth]{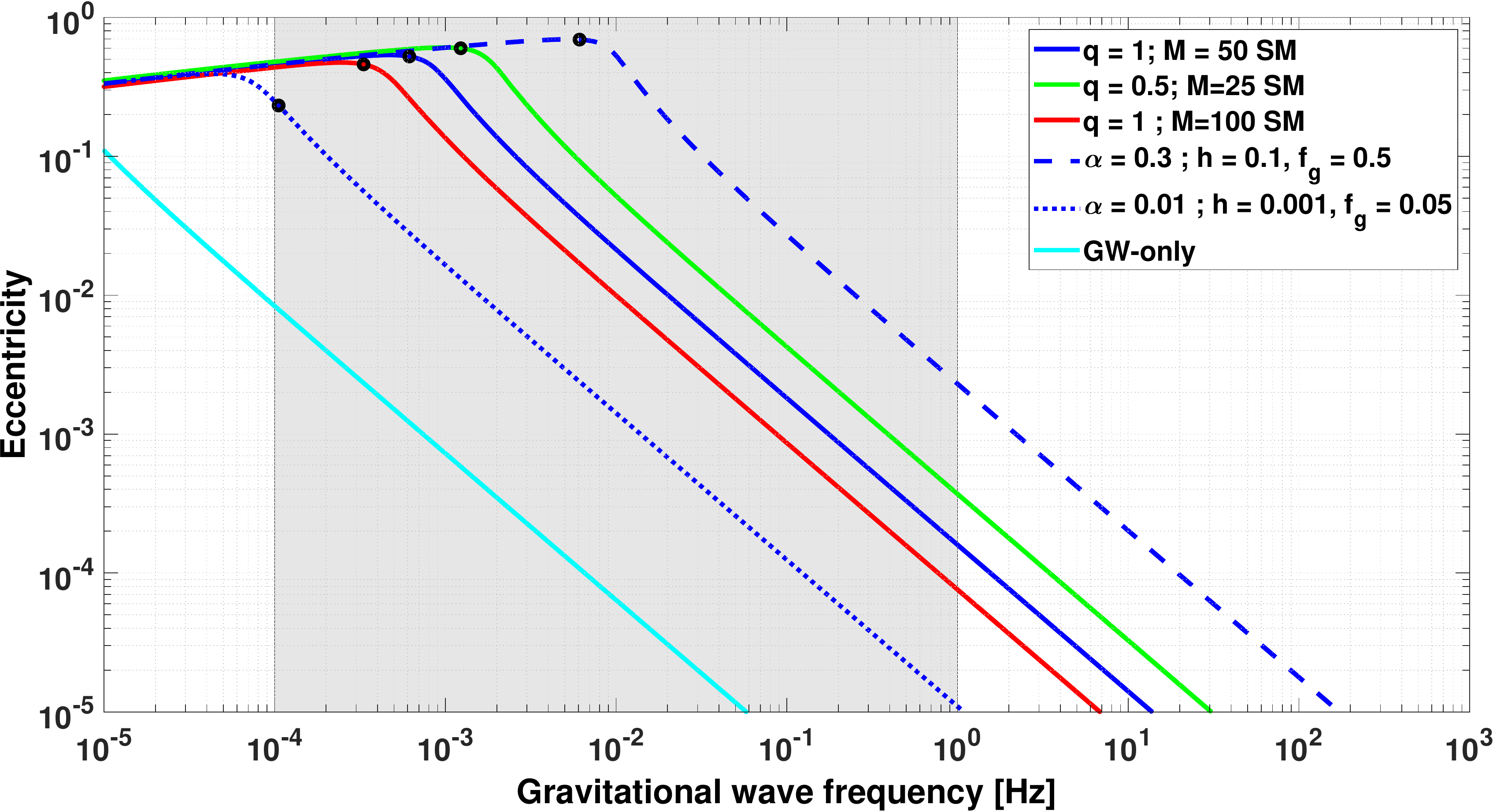}
    \caption{Orbital eccentricity as a function of peak gravitational wave frequency. The grey shaded area represents the frequency range observable by LISA.  The cyan line displays the GW-only evolution. The other curves include the disc-binary interaction where orbital parameters (binary mass $M$ and mass ratio $q$) and disc parameters (viscosity parameter $\alpha$, aspect ratio $h,$ and gas fraction $f_g$) are varied; see the figure caption for the specific values.}
    \label{e_f}
\end{figure}

Figure $\ref{e_f}$ shows the orbital eccentricity as a function of the GW frequency for different binary and disc configurations. We clearly see two distinct trends: the eccentricity growth in the disc-driven regime at lower $f_\text{GW}$, and the eccentricity decay in the GW-driven regime at higher $f_\text{GW}$. It has been argued that LISA should be able to detect a non-zero eccentricity for all binaries with $e \gtrsim 10^{-2}$, and for a $\sim 90\%$ fraction of binaries with $e \gtrsim 10^{-3}$ (assuming an observation time of 5yr) \citep{Nishizawa_et_2016}. From Fig. \ref{e_f} we observe that the majority of binaries evolving in AGN accretion discs should have detectable eccentricities in the LISA band, with typical values in the range $e \sim (0.01-0.1)$ at $f_\text{GW} = 10^{-2}$ Hz. 

By not taking into account the effects of the disc-binary interaction,  the orbital evolution is purely dictated by GW-emission. In our channel, this would mean that the eccentricity growth from the disc-binary interaction is absent. In this case, the binaries would enter the LISA band at a stage when they possibly have eccentricities that are too low for detection (unless the initial eccentricity is as high as $e_0 \sim 0.9$). 
Compared to a purely GW-driven evolution, the eccentricity growth in the disc-driven regime tends to shift the binaries towards frequencies more easily detectable by LISA (cf. Eq. \ref{Eq_e_f}). Therefore an intermediate disc-driven phase is essential in order to obtain potentially measurable eccentricities in the LISA band. By the time the binaries enter the LIGO frequency band ($\sim 10$ Hz), they are all likely to be completely circularised and hence have no measurable eccentricities. 

The overall eccentricity evolution is determined by both the binary orbit and accretion disc parameters. 
We recall here the dependence of the eccentricity growth/decay in the disc/GW regime on the total binary mass and mass ratio: $\dot{e}_\text{disc} \propto M_b^{-3/2} \frac{(1+q)^2}{q}$ and $\dot{e}_\text{GW} \propto - M_b^3 \frac{q}{(1+q)^2}$ (Eqs. \ref {Eq_disc_analy_e}-\ref{Eq_GW_analy_e}).  
For a larger binary mass, the rate of eccentricity growth is smaller in the disc-regime, while the rate of eccentricity decay is larger in the GW-regime. The two opposite dependences on $M_b$ combine to give a smaller resulting eccentricity  at the later stages of this evolution channel.  Likewise, for a larger mass ratio, the rate of eccentricity growth is smaller in the disc-regime, while the rate of eccentricity decay is larger in the GW-regime, and the resulting eccentricity is smaller. By combining the two dependences, massive binaries with large mass ratios should have the smallest eccentricities. Conversely, less massive binaries with smaller mass ratios should have the highest eccentricities. 
These trends are graphically illustrated in Fig. \ref{e_f}: at a given GW frequency, the less massive binary with the smaller mass ratio has a higher eccentricity compared to the fiducial case, while the opposite is true for the more massive binary. Therefore if we are interested in detecting high residual eccentricities, then according to the here-presented evolution channel, we should look for less massive, unequal-mass binaries.

The important phase of the eccentricity growth is mainly governed by the accretion disc parameters.  
We recall that the rate of eccentricity growth in the disc-driven regime scales as $\dot{e}_\text{disc} \propto \alpha f_g h^2$. Thus an increase in the viscosity parameter, gas fraction, and disc aspect ratio lead to more efficient eccentricity growth in the disc-regime, yielding a higher residual eccentricity. In Fig. \ref{e_f} we see that the largest combination of the three accretion disc parameters (large $\alpha, f_g, h$) leads to the highest residual eccentricities, with values of $e \sim 0.5$ at $f_\text{GW} = 10^{-2}$ Hz. On the contrary, the smallest combination of the three parameters leads to the lowest eccentricities, with values of $e \sim 10^{-3}$ at $f_\text{GW} = 10^{-2}$ Hz. As a consequence, the global range of the predicted eccentricity evolution seems to be primarily determined by the AGN accretion disc parameters. 


\section{Discussion}
\label{Sect_discussion}

Among the different BBH formation scenarios discussed in the literature, a promising channel is the evolution of BBHs in AGN accretion discs. In this picture, gaseous torques due to the disc-binary interaction may help to foster the orbital decay into the domain of GW inspiral. In this paper we investigate coupled disc+GW-driven evolution using a simple analytic model that takes into account various orbital and disc parameters. 
Below we comment on the characteristic properties (merger timescale and eccentricity evolution) of this particular evolution channel, also in relation to other works. We then discuss the limitations of our model regarding accretion flows and cavity shapes, and briefly mention possible directions for future research. 


\subsection{Characteristics of the coupled disc+GW-driven evolution scenario}
\label{Subsect_characteristics}

We have followed the dynamics of BBHs as they transition from the disc-driven regime into the GW-driven regime (Sect. $\ref{Sect_bin_evol}$). In our model setup, two characteristic features of the disc-dominated phase help to bring the binary into the GW-domain: the decay of the semi-major axis and the growth of the orbital eccentricity, both induced by the disc-binary interaction. 
Early SPH simulations already showed that the semi-major axis decreases, while at the same time the orbital eccentricity increases \citep[][]{Artymowicz_et_1991}, a phenomenon verified by further simulations. For example, the global trend is supported by numerical simulations of SMBH binaries in gas-rich discs, which display simultaneous eccentricity growth and orbital decay \citep{Armitage_Natarajan_2005}. 
Subsequent simulations of the dynamics of SMBH binaries embedded in self-gravitating discs confirm that the semi-major axis decreases, while the binary eccentricity increases, as a result of the disc-binary coupling \citep{Cuadra_et_2009}. N-body SPH simulations of gaseous protoplanetary discs suggest that eccentric binaries undergo significant eccentricity growth, whereas quasi-circular binaries exhibit no appreciable eccentricity excitation \citep{Fleming_Quinn_2017}. 
While the overall trend of orbital decay and eccentricity growth seems to be reproduced by various numerical simulations, the situation can change for strongly accreting binaries, as suggested by recent hydrodynamical simulations (see Sect. $\ref{Subsect_accretion}$). 

Different studies broadly agree that BBH mergers could be fostered in gaseous environments, but the efficiency of such gas-assisted orbital decay remains uncertain. 
In most previous work, the average merger time for binaries embedded in AGN accretion discs was not explicitly computed, and merger rates were for example parametrised through the AGN disc lifetimes \citep[e.g][]{McKernan_et_2018}. 
Most recently, \citet{Tagawa_et_2019} performed 1D N-body simulations to self-consistently follow the evolution of binaries in AGN discs, from their formation to disruption. The main BBH formation channel is attributed to gas capture within the AGN disc, while binaries are disrupted by soft binary-single interactions. Several physical processes, including gas dynamical friction, type I/II migration torques, GW emission, and multi-body stellar interactions, are implemented via semi-analytical prescriptions. Only circular binaries are considered in their model setup, and a potential orbital eccentricity evolution is ignored. The resulting BBH merger rate in AGN discs lies in the range $\mathcal{R} = (0.02 - 60) \, \, \mathrm{Gpc^{-3} yr^{-1}}$ \citep{Tagawa_et_2019}. 
Their merger rate is parametrised through the merger fraction per AGN lifetime: $\mathcal{R} \propto f_\textrm{BH, merge}/t_\textrm{AGN}$, where $t_\textrm{AGN}$ is the average lifetime of AGN discs. However, the actual AGN lifetime is a notoriously difficult quantity to constrain, also considering the variable nature of the central AGN  \citep{Martini_2004, Hickox_et_2014}.

With our simple analytic model we can provide a quantification of the merger time resulting from the coupled disc+GW-driven evolution. In our picture, the merger timescale $\tau_\text{merger}$ is directly obtained from the numerical integration of the coupled evolution equations (Eqs. \ref{Eq_a_coupled} - \ref{Eq_e_coupled}). We analyse its dependence on both the accretion disc and binary orbital parameters (Sect. \ref{Sect_mergertime}). We show that the merger timescale can be reduced by several orders of magnitude when including the disc-binary interaction compared to a purely GW-driven evolution. From our parameter space study, we obtain that the merger timescale is shortest for binaries embedded in AGN accretion discs with a large viscosity parameter, large gas fraction, and large disc aspect ratio. Moreover, the eccentricity growth in the disc-regime allows the binary to enter the GW-regime with high eccentricity, such that the subsequent evolution can be further speeded up (due to the steep $e$-dependence in the GW regime).

A key element that may help discriminate between different BBH paths is the evolution of the orbital eccentricity, in particular in the LISA frequency band \citep[e.g.][]{Armitage_Natarajan_2005}. 
The eccentricity evolution in the case of GW-only decay has been previously considered, for instance when comparing the two main BBH channels: isolated formation in galactic fields versus dynamical formation in globular clusters \citep{Breivik_et_2016, Dorazio_Samsing_2018}. A novel result here is that we follow the eccentricity evolution in the coupled disc+GW evolution channel, from the $e$-growth in the disc regime to the $e$-decay in the GW regime (Sect. \ref{Sect_ecc_evol}). In this picture, due to the eccentricity growth in the disc-driven regime, we expect significant residual eccentricities at LISA frequencies. This trend is also in qualitative agreement with the SPH simulations of SMBH binaries embedded in CBDs \citep{Roedig_et_2011}. 
The temporary eccentricity growth in the disc regime (only partly damped in the subsequent GW regime) ensures that the binaries retain significant residual eccentricities, potentially detectable by LISA.  
This highlights again the importance of the intermediate disc-driven regime in the coupled disc+GW evolution scenario. 
Reversing the argument, high residual eccentricities observed in the LISA band could also be suggestive of BBH formation in AGN accretion discs. 
By computing the BBH merger rate density that results from the coupled disc+GW evolution channel, we obtain typical rates in the range $\mathcal{R} \sim (0.002 - 18) \, \mathrm{Gpc^{-3} yr^{-1}}$ \citep{Paper_GWRates}. 


\subsection{Role of gas inflows and accretion streams } 
\label{Subsect_accretion}

In our idealised analytic model, we assume that the binary resides in a depleted cavity, and we neglect gas inflows and subsequent accretion onto the individual BHs. 
In reality, gas streams can penetrate within the cavity, and mass transfer does occur from the CBD to the binary. 
A variety of numerical simulations show the development of narrow gas streams flowing from the inner edge of the disc towards the binary, eventually leading to the formation of mini-discs surrounding each BH \citep{MacFadyen_Milosavljevic_2008, Shi_et_2012, DOrazio_et_2013, Farris_et_2014, Tang_et_2017, Miranda_et_2017, Moody_et_2019, Munoz_et_2019, Duffell_et_2019}. 

Those numerical studies indicate that the mass accretion rate onto the binary can be considerable. Two-dimensional hydrodynamical simulations show significant mass transfer from the CBD onto equal-mass binaries, albeit with a somewhat reduced accretion rate compared to the case without binary torques \citep{MacFadyen_Milosavljevic_2008}. A follow-up study extending the analysis to different mass ratios suggests that the time-averaged accretion rate can be comparable to that on a single central mass \citep{DOrazio_et_2013}. 
Three-dimensional MHD simulations confirm that the time-averaged accretion rate is essentially indistinguishable from that on a single BH \citep{Shi_Krolik_2015}. Two-dimensional hydrodynamical simulations including the inner cavity (without the excision of the innermost central region as done in previous works) further indicate that the accretion rate is efficiently channelled into narrow streams fuelling the mini-discs \citep{Farris_et_2014}. 

Accretion flows within the cavity can have a major impact on the orbital evolution of the binary. Angular momentum is carried with the accreting matter, and such advection needs to be accounted for in the torque balance. In general, the binary loses angular momentum to the disc though gravitational torques, whereas it gains angular momentum through accretion torques. The orbital evolution is thus governed by the balance between the two competing mechanisms. The large accretion rates observed in 3D MHD simulations suggest a near-cancellation of the opposite torques, but still imply a small shrinkage rate with negative $\dot{a}/a < 0$ \citep{Shi_et_2012}. The torques are also found to depend on the sink prescription in 2D viscous hydrodynamical simulations: slower sinks result in negative torques, while the torques may become positive for faster sinks \citep{Tang_et_2017}. Long-duration 2D simulations of eccentric binaries suggest that the net angular momentum received by the binary is mostly positive, such that the binary separation grows in time \citep{Miranda_et_2017}. 

More recent 2D hydrodynamical simulations (which explicitly compute the gas dynamics within the cavity) provide evidence that equal-mass eccentric binaries can consistently gain net angular momentum \citep{Munoz_et_2019}. As a consequence, the accreting binaries tend to expand rather than shrink.  
This result seems to hold for moderate mass ratios, and also when considering a finite mass supply \citep{Munoz_et_2020}. 
The overall sign of the net torque (positive $\langle \dot{a} \rangle > 0$) is also corroborated in the first 3D hydrodynamic simulations: accreting equal-mass binaries on circular orbits expand in both aligned and misaligned discs \citep{Moody_et_2019}. 
Most recently, analysing variations of the mass ratio ($q = 0.01-1$) and disc viscosity ($\alpha = 0.03-0.15$) in 2D simulations, \citet{Duffell_et_2019} obtain that the net torque is always positive for $q \gtrsim 0.05$, with little dependence on the viscosity parameter. 

The aforementioned numerical simulations suggest that the binary separation increases with time, implying that binaries undergo expansion instead of contraction. Therefore our overall picture of gas-aided orbital inspiral could potentially be reversed, such that accreting binaries may not reach the merger stage. By completely neglecting the mass transfer from the CBD, we may over-estimate the orbital decay and eccentricity growth in our simplified analytic model. 
Nonetheless, our model results may be most relevant in regimes of low accretion rates \citep[cf.][]{Hayasaki_2009}.  

In this context, we note that most of the previously discussed 2D and 3D simulations assume disc aspect ratios of the order of $h \sim 0.1$, due to computational limitations. However, a lower value of $h \sim (10^{-2}-10^{-3})$ should be more appropriate for the thin discs in AGNs \citep[][and references therein]{Ragusa_et_2016}. For aspect ratios of $h \lesssim 10^{-2}$ (typical of AGN discs), it has been argued that accretion flows can be significantly suppressed.
In particular, 3D SPH simulations show that the mass accretion rate is considerably reduced in such thin discs, with only a fraction of the gas actually flowing into the cavity and accreting onto the binary (following a roughly linear decrease with the aspect ratio for $h < 0.1$) \citep{Ragusa_et_2016}. A similar suppression of the mass accretion rate for thin discs is also observed in 2D hydrodynamic simulations \citep{Terquem_Papaloizou_2017}. Interestingly, such a dependence on the disc thickness was already pointed out by \citet{Artymowicz_Lubow_1996}. In conclusion, the details of the disc-binary interaction are sensitive to the disc aspect ratio and a more thorough analysis of this important parameter is required. 

Until recently, most numerical studies have focused on particular binary configurations, such as near-equal-mass binaries \citep{MacFadyen_Milosavljevic_2008, Shi_et_2012, Munoz_et_2019, Moody_et_2019} and/or quasi-circular orbits \citep{Shi_et_2012, Tang_et_2017, Moody_et_2019, Munoz_et_2020}. However, the actual torque balance depends on a number of different factors, including the binary mass ratio $q$ and orbital eccentricity $e$, as well as the disc thickness $H$ and disc viscosity $\alpha$. For instance, \citet{Duffell_et_2019} hint at the eccentricity as being a possible parameter that could alter the binary orbital trend. In fact, the magnitude of the expansion rate tends to be smaller for eccentric binaries ($e \sim 0.6$) compared to circular counterparts \citep{Munoz_et_2019}. The key role of the orbital eccentricity is emphasised in our coupled disc+GW evolution scenario (Sects. $\ref{Sect_bin_evol}$-$\ref{Sect_ecc_evol}$), and the uncertain $e$-evolution due to the complex disc-binary coupling deserves further investigation. 

Finally, the BBH-in-AGN evolution channel has another unique physical implication related to the accretion streams: this is the only scenario in which we can plausibly expect electromagnetic (EM) counterparts from the merger of two stellar-mass black holes.
In fact, BBHs located in the dense gaseous environment of AGN accretion discs may accrete at super-Eddington rates \citep{Stone_et_2017, Bartos_et_2017}. As a result, they may give rise to EM radiation over a broad wavelength range, from radio waves up to gamma-rays \citep{Murase_et_2016, Perna_et_2019}.  
By properly modelling the accretion flows through the CBD and mini-discs, one can possibly put some physical constraints on the resulting EM radiation that could be detected for example by space-based observatories \citep{Bartos_et_2017, Ford_et_2019}. 
Furthermore, several numerical simulations suggest a time-variable luminosity output, with the light curves following characteristic patterns (for example quasi-periodic oscillations modulated by the binary and/or disc orbits) that could be useful in identifying potential EM counterparts \citep{MacFadyen_Milosavljevic_2008, DOrazio_et_2013, Moody_et_2019, Duffell_et_2019, Munoz_et_2020}. 
In future studies, we aim to refine our modelling by properly including the accretion process to obtain a more complete picture of both binary dynamics and potential EM counterparts within the BBH-in-AGN scenario. 


\begin{acknowledgements}
WI acknowledges support from the University of Zurich.
\end{acknowledgements}


\bibliographystyle{aa}
\bibliography{biblio}

\begin{thebibliography}{55}
\expandafter\ifx\csname natexlab\endcsname\relax\def\natexlab#1{#1}\fi

\bibitem[{{Abbott} {et~al.}(2019){Abbott}, {Abbott}, {Abbott}, {Abraham},
  {Acernese}, {Ackley}, {Adams}, {Adhikari}, {Adya}, {Affeldt}, {Agathos},
  {Agatsuma}, {Aggarwal}, {Aguiar}, {Aiello}, {Ain}, {Ajith}, {Allen},
  {Allocca}, {Aloy}, {Altin}, {Amato}, {Ananyeva}, {Anderson}, {Anderson},
  {Angelova}, {Antier}, {Appert}, {Arai}, {Araya}, {Areeda}, {Ar{\`e}ne},
  {Arnaud}, {Arun}, {Ascenzi}, {Ashton}, {Aston}, {Astone}, {Aubin}, {Aufmuth},
  {AultONeal}, {Austin}, {Avendano}, {Avila-Alvarez}, {Babak}, {Bacon},
  {Badaracco}, {Bader}, {Bae}, {Baker}, {Baldaccini}, {Ballardin}, {Ballmer},
  {Banagiri}, {Barayoga}, {Barclay}, {Barish}, {Barker}, {Barkett}, {Barnum},
  {Barone}, {Barr}, {Barsotti}, {Barsuglia}, {Barta}, {Bartlett}, {Bartos},
  {Bassiri}, {Basti}, {Bawaj}, {Bayley}, {Bazzan}, {B{\'e}csy}, {Bejger},
  {Belahcene}, {Bell}, {Beniwal}, {Berger}, {Bergmann}, {Bernuzzi}, {Bero},
  {Berry}, {Bersanetti}, {Bertolini}, {Betzwieser}, {Bhand are}, {Bidler},
  {Bilenko}, {Bilgili}, {Billingsley}, {Birch}, {Birney}, {Birnholtz},
  {Biscans}, {Biscoveanu}, {Bisht}, {Bitossi}, {Bizouard}, {Blackburn},
  {Blair}, {Blair}, {Blair}, {Bloemen}, {Bode}, {Boer}, {Boetzel}, {Bogaert},
  {Bondu}, {Bonilla}, {Bonnand}, {Booker}, {Boom}, {Booth}, {Bork}, {Boschi},
  {Bose}, {Bossie}, {Bossilkov}, {Bosveld}, {Bouffanais}, {Bozzi},
  {Bradaschia}, {Brady}, {Bramley}, {Branchesi}, {Brau}, {Briant}, {Briggs},
  {Brighenti}, {Brillet}, {Brinkmann}, {Brisson}, {Brockill}, {Brooks},
  {Brown}, {Brunett}, {Buikema}, {Bulik}, {Bulten}, {Buonanno}, {Buscicchio},
  {Buskulic}, {Buy}, {Byer}, {Cabero}, {Cadonati}, {Cagnoli}, {Cahillane},
  {Calder{\'o}n Bustillo}, {Callister}, {Calloni}, {Camp}, {Campbell},
  {Canepa}, {Cannon}, {Cao}, {Cao}, {Capocasa}, {Carbognani}, {Caride},
  {Carney}, {Carullo}, {Casanueva Diaz}, {Casentini}, {Caudill},
  {Cavagli{\`a}}, {Cavalier}, {Cavalieri}, {Cella}, {Cerd{\'a}-Dur{\'a}n},
  {Cerretani}, {Cesarini}, {Chaibi}, {Chakravarti}, {Chamberlin}, {Chan},
  {Chao}, {Charlton}, {Chase}, {Chassand e-Mottin}, {Chatterjee}, {Chaturvedi},
  {Chatziioannou}, {Cheeseboro}, {Chen}, {Chen}, {Chen}, {Cheng}, {Cheong},
  {Chia}, {Chincarini}, {Chiummo}, {Cho}, {Cho}, {Cho}, {Christensen}, {Chu},
  {Chua}, {Chung}, {Chung}, {Ciani}, {Ciobanu}, {Ciolfi}, {Cipriano}, {Cirone},
  {Clara}, {Clark}, {Clearwater}, {Cleva}, {Cocchieri}, {Coccia}, {Cohadon},
  {Cohen}, {Colgan}, {Colleoni}, {Collette}, {Collins}, {Cominsky},
  {Constancio}, {Conti}, {Cooper}, {Corban}, {Corbitt}, {Cordero-Carri{\'o}n},
  {Corley}, {Cornish}, {Corsi}, {Cortese}, {Costa}, {Cotesta}, {Coughlin},
  {Coughlin}, {Coulon}, {Countryman}, {Couvares}, {Covas}, {Cowan}, {Coward},
  {Cowart}, {Coyne}, {Coyne}, {Creighton}, {Creighton}, {Cripe}, {Croquette},
  {Crowder}, {Cullen}, {Cumming}, {Cunningham}, {Cuoco}, {Dal Canton},
  {D{\'a}lya}, {Danilishin}, {D'Antonio}, {Danzmann}, {Dasgupta}, {Da Silva
  Costa}, {Datrier}, {Dattilo}, {Dave}, {Davier}, {Davis}, {Daw}, {DeBra},
  {Deenadayalan}, {Degallaix}, {De Laurentis}, {Del{\'e}glise}, {Del Pozzo},
  {DeMarchi}, {Demos}, {Dent}, {De Pietri}, {Derby}, {De Rosa}, {De Rossi},
  {DeSalvo}, {de Varona}, {Dhurandhar}, {D{\'\i}az}, {Dietrich}, {Di Fiore},
  {Di Giovanni}, {Di Girolamo}, {Di Lieto}, {Ding}, {Di Pace}, {Di Palma}, {Di
  Renzo}, {Dmitriev}, {Doctor}, {Donovan}, {Dooley}, {Doravari}, {Dorrington},
  {Downes}, {Drago}, {Driggers}, {Du}, {Ducoin}, {Dupej}, {Dwyer}, {Easter},
  {Edo}, {Edwards}, {Effler}, {Ehrens}, {Eichholz}, {Eikenberry}, {Eisenmann},
  {Eisenstein}, {Essick}, {Estelles}, {Estevez}, {Etienne}, {Etzel}, {Evans},
  {Evans}, {Fafone}, {Fair}, {Fairhurst}, {Fan}, {Farinon}, {Farr}, {Farr},
  {Fauchon-Jones}, {Favata}, {Fays}, {Fazio}, {Fee}, {Feicht}, {Fejer}, {Feng},
  {Fernand ez-Galiana}, {Ferrante}, {Ferreira}, {Ferreira}, {Ferrini},
  {Fidecaro}, {Fiori}, {Fiorucci}, {Fishbach}, {Fisher}, {Fishner},
  {Fitz-Axen}, {Flaminio}, {Fletcher}, {Flynn}, {Fong}, {Font}, {Forsyth},
  {Fournier}, {Frasca}, {Frasconi}, {Frei}, {Freise}, {Frey}, {Frey},
  {Fritschel}, {Frolov}, {Fulda}, {Fyffe}, {Gabbard}, {Gadre}, {Gaebel},
  {Gair}, {Gammaitoni}, {Ganija}, {Gaonkar}, {Garcia},
  {Garc{\'\i}a-Quir{\'o}s}, {Garufi}, {Gateley}, {Gaudio}, {Gaur}, {Gayathri},
  {Gemme}, {Genin}, {Gennai}, {George}, {George}, {Gergely}, {Germain},
  {Ghonge}, {Ghosh}, {Ghosh}, {Ghosh}, {Giacomazzo}, {Giaime}, {Giardina},
  {Giazotto}, {Gill}, {Giordano}, {Glover}, {Godwin}, {Goetz}, {Goetz},
  {Goncharov}, {Gonz{\'a}lez}, {Gonzalez Castro}, {Gopakumar}, {Gorodetsky},
  {Gossan}, {Gosselin}, {Gouaty}, {Grado}, {Graef}, {Granata}, {Grant}, {Gras},
  {Grassia}, {Gray}, {Gray}, {Greco}, {Green}, {Green}, {Gretarsson}, {Groot},
  {Grote}, {Grunewald}, {Gruning}, {Guidi}, {Gulati}, {Guo}, {Gupta}, {Gupta},
  {Gustafson}, {Gustafson}, {Haegel}, {Halim}, {Hall}, {Hall}, {Hamilton},
  {Hammond}, {Haney}, {Hanke}, {Hanks}, {Hanna}, {Hannam}, {Hannuksela},
  {Hanson}, {Hardwick}, {Haris}, {Harms}, {Harry}, {Harry}, {Haster},
  {Haughian}, {Hayes}, {Healy}, {Heidmann}, {Heintze}, {Heitmann}, {Hello},
  {Hemming}, {Hendry}, {Heng}, {Hennig}, {Heptonstall}, {Hernandez Vivanco},
  {Heurs}, {Hild}, {Hinderer}, {Hoak}, {Hochheim}, {Hofman}, {Holgado},
  {Holland }, {Holt}, {Holz}, {Hopkins}, {Horst}, {Hough}, {Howell}, {Hoy},
  {Hreibi}, {Huerta}, {Huet}, {Hughey}, {Hulko}, {Husa}, {Huttner},
  {Huynh-Dinh}, {Idzkowski}, {Iess}, {Ingram}, {Inta}, {Intini}, {Irwin},
  {Isa}, {Isac}, {Isi}, {Iyer}, {Izumi}, {Jacqmin}, {Jadhav}, {Jani},
  {Janthalur}, {Jaranowski}, {Jenkins}, {Jiang}, {Johnson}, {Jones}, {Jones},
  {Jones}, {Jonker}, {Ju}, {Junker}, {Kalaghatgi}, {Kalogera}, {Kamai}, {Kand
  hasamy}, {Kang}, {Kanner}, {Kapadia}, {Karki}, {Karvinen}, {Kashyap},
  {Kasprzack}, {Katsanevas}, {Katsavounidis}, {Katzman}, {Kaufer}, {Kawabe},
  {Keerthana}, {K{\'e}f{\'e}lian}, {Keitel}, {Kennedy}, {Key}, {Khalili},
  {Khan}, {Khan}, {Khan}, {Khan}, {Khazanov}, {Khursheed}, {Kijbunchoo}, {Kim},
  {Kim}, {Kim}, {Kim}, {Kim}, {Kim}, {Kimball}, {King}, {King},
  {Kinley-Hanlon}, {Kirchhoff}, {Kissel}, {Kleybolte}, {Klika}, {Klimenko},
  {Knowles}, {Koch}, {Koehlenbeck}, {Koekoek}, {Koley}, {Kondrashov}, {Kontos},
  {Koper}, {Korobko}, {Korth}, {Kowalska}, {Kozak}, {Kringel}, {Krishnendu},
  {Kr{\'o}lak}, {Kuehn}, {Kumar}, {Kumar}, {Kumar}, {Kumar}, {Kuo}, {Kutynia},
  {Kwang}, {Lackey}, {Lai}, {Lam}, {Landry}, {Lane}, {Lang}, {Lange}, {Lantz},
  {Lanza}, {Lartaux-Vollard}, {Lasky}, {Laxen}, {Lazzarini}, {Lazzaro},
  {Leaci}, {Leavey}, {Lecoeuche}, {Lee}, {Lee}, {Lee}, {Lee}, {Lee}, {Lee},
  {Lehmann}, {Lenon}, {Leroy}, {Letendre}, {Levin}, {Li}, {Li}, {Li}, {Li},
  {Lin}, {Linde}, {Linker}, {Littenberg}, {Liu}, {Liu}, {Lo}, {Lockerbie},
  {London}, {Longo}, {Lorenzini}, {Loriette}, {Lormand}, {Losurdo}, {Lough},
  {Lousto}, {Lovelace}, {Lower}, {L{\"u}ck}, {Lumaca}, {Lundgren}, {Lynch},
  {Ma}, {Macas}, {Macfoy}, {MacInnis}, {Macleod}, {Macquet},
  {Maga{\~n}a-Sandoval}, {Maga{\~n}a Zertuche}, {Magee}, {Majorana},
  {Maksimovic}, {Malik}, {Man}, {Mandic}, {Mangano}, {Mansell}, {Manske},
  {Mantovani}, {Mapelli}, {Marchesoni}, {Marion}, {M{\'a}rka}, {M{\'a}rka},
  {Markakis}, {Markosyan}, {Markowitz}, {Maros}, {Marquina}, {Marsat},
  {Martelli}, {Martin}, {Martin}, {Martynov}, {Mason}, {Massera}, {Masserot},
  {Massinger}, {Masso-Reid}, {Mastrogiovanni}, {Matas}, {Matichard}, {Matone},
  {Mavalvala}, {Mazumder}, {McCann}, {McCarthy}, {McClelland }, {McCormick},
  {McCuller}, {McGuire}, {McIver}, {McManus}, {McRae}, {McWilliams}, {Meacher},
  {Meadors}, {Mehmet}, {Mehta}, {Meidam}, {Melatos}, {Mendell}, {Mercer},
  {Mereni}, {Merilh}, {Merzougui}, {Meshkov}, {Messenger}, {Messick},
  {Metzdorff}, {Meyers}, {Miao}, {Michel}, {Middleton}, {Mikhailov}, {Milano},
  {Miller}, {Miller}, {Millhouse}, {Mills}, {Milovich-Goff}, {Minazzoli},
  {Minenkov}, {Mishkin}, {Mishra}, {Mistry}, {Mitra}, {Mitrofanov},
  {Mitselmakher}, {Mittleman}, {Mo}, {Moffa}, {Mogushi}, {Mohapatra},
  {Montani}, {Moore}, {Moraru}, {Moreno}, {Morisaki}, {Mours}, {Mow-Lowry},
  {Mukherjee}, {Mukherjee}, {Mukherjee}, {Mukund}, {Mullavey}, {Munch},
  {Mu{\~n}iz}, {Muratore}, {Murray}, {Nagar}, {Nardecchia}, {Naticchioni},
  {Nayak}, {Neilson}, {Nelemans}, {Nelson}, {Nery}, {Neunzert}, {Ng}, {Ng},
  {Nguyen}, {Nichols}, {Nissanke}, {Nocera}, {North}, {Nuttall},
  {Obergaulinger}, {Oberling}, {O'Brien}, {O'Dea}, {Ogin}, {Oh}, {Oh}, {Ohme},
  {Ohta}, {Okada}, {Oliver}, {Oppermann}, {Oram}, {O'Reilly}, {Ormiston},
  {Ortega}, {O'Shaughnessy}, {Ossokine}, {Ottaway}, {Overmier}, {Owen}, {Pace},
  {Pagano}, {Page}, {Pai}, {Pai}, {Palamos}, {Palashov}, {Palomba},
  {Pal-Singh}, {Pan}, {Pang}, {Pang}, {Pankow}, {Pannarale}, {Pant},
  {Paoletti}, {Paoli}, {Parida}, {Parker}, {Pascucci}, {Pasqualetti},
  {Passaquieti}, {Passuello}, {Patil}, {Patricelli}, {Pearlstone}, {Pedersen},
  {Pedraza}, {Pedurand}, {Pele}, {Penn}, {Perez}, {Perreca}, {Pfeiffer},
  {Phelps}, {Phukon}, {Piccinni}, {Pichot}, {Piergiovanni}, {Pillant},
  {Pinard}, {Pirello}, {Pitkin}, {Poggiani}, {Pong}, {Ponrathnam}, {Popolizio},
  {Porter}, {Powell}, {Prajapati}, {Prasad}, {Prasai}, {Prasanna}, {Pratten},
  {Prestegard}, {Privitera}, {Prodi}, {Prokhorov}, {Puncken}, {Punturo},
  {Puppo}, {P{\"u}rrer}, {Qi}, {Quetschke}, {Quinonez}, {Quintero},
  {Quitzow-James}, {Raab}, {Radkins}, {Radulescu}, {Raffai}, {Raja}, {Rajan},
  {Rajbhandari}, {Rakhmanov}, {Ramirez}, {Ramos-Buades}, {Rana}, {Rao},
  {Rapagnani}, {Raymond}, {Razzano}, {Read}, {Regimbau}, {Rei}, {Reid},
  {Reitze}, {Ren}, {Ricci}, {Richardson}, {Richardson}, {Ricker}, {Riles},
  {Rizzo}, {Robertson}, {Robie}, {Robinet}, {Rocchi}, {Rolland}, {Rollins},
  {Roma}, {Romanelli}, {Romano}, {Romel}, {Romie}, {Rose}, {Rosi{\'n}ska},
  {Rosofsky}, {Ross}, {Rowan}, {R{\"u}diger}, {Ruggi}, {Rutins}, {Ryan},
  {Sachdev}, {Sadecki}, {Sakellariadou}, {Salconi}, {Saleem}, {Samajdar},
  {Sammut}, {Sanchez}, {Sanchez}, {Sanchis-Gual}, {Sandberg}, {Sand ers},
  {Santiago}, {Sarin}, {Sassolas}, {Sathyaprakash}, {Saulson}, {Sauter},
  {Savage}, {Schale}, {Scheel}, {Scheuer}, {Schmidt}, {Schnabel}, {Schofield},
  {Sch{\"o}nbeck}, {Schreiber}, {Schulte}, {Schutz}, {Schwalbe}, {Scott},
  {Scott}, {Seidel}, {Sellers}, {Sengupta}, {Sennett}, {Sentenac}, {Sequino},
  {Sergeev}, {Setyawati}, {Shaddock}, {Shaffer}, {Shahriar}, {Shaner}, {Shao},
  {Sharma}, {Shawhan}, {Shen}, {Shink}, {Shoemaker}, {Shoemaker},
  {ShyamSundar}, {Siellez}, {Sieniawska}, {Sigg}, {Silva}, {Singer}, {Singh},
  {Singhal}, {Sintes}, {Sitmukhambetov}, {Skliris}, {Slagmolen},
  {Slaven-Blair}, {Smith}, {Smith}, {Somala}, {Son}, {Sorazu}, {Sorrentino},
  {Souradeep}, {Sowell}, {Spencer}, {Spera}, {Srivastava}, {Srivastava},
  {Staats}, {Stachie}, {Standke}, {Steer}, {Steinke}, {Steinlechner},
  {Steinlechner}, {Steinmeyer}, {Stevenson}, {Stocks}, {Stone}, {Stops},
  {Strain}, {Stratta}, {Strigin}, {Strunk}, {Sturani}, {Stuver}, {Sudhir},
  {Summerscales}, {Sun}, {Sunil}, {Suresh}, {Sutton}, {Swinkels},
  {Szczepa{\'n}czyk}, {Tacca}, {Tait}, {Talbot}, {Talukder}, {Tanner},
  {T{\'a}pai}, {Taracchini}, {Tasson}, {Taylor}, {Thies}, {Thomas}, {Thomas},
  {Thondapu}, {Thorne}, {Thrane}, {Tiwari}, {Tiwari}, {Tiwari}, {Toland},
  {Tonelli}, {Tornasi}, {Torres-Forn{\'e}}, {Torrie}, {T{\"o}yr{\"a}},
  {Travasso}, {Traylor}, {Tringali}, {Trovato}, {Trozzo}, {Trudeau}, {Tsang},
  {Tse}, {Tso}, {Tsukada}, {Tsuna}, {Tuyenbayev}, {Ueno}, {Ugolini},
  {Unnikrishnan}, {Urban}, {Usman}, , {LIGO Scientific Collaboration}, \&
  {Virgo Collaboration}}]{Abbott_et_2019_runs}
{Abbott}, B.~P., {Abbott}, R., {Abbott}, T.~D., {et~al.} 2019, \apjl, 882, L24

\bibitem[{{Armitage} \& {Natarajan}(2005)}]{Armitage_Natarajan_2005}
{Armitage}, P.~J. \& {Natarajan}, P. 2005, \apj, 634, 921

\bibitem[{{Artymowicz} {et~al.}(1991){Artymowicz}, {Clarke}, {Lubow}, \&
  {Pringle}}]{Artymowicz_et_1991}
{Artymowicz}, P., {Clarke}, C.~J., {Lubow}, S.~H., \& {Pringle}, J.~E. 1991,
  \apjl, 370, L35

\bibitem[{{Artymowicz} \& {Lubow}(1994)}]{Artymowicz_Lubow_1994}
{Artymowicz}, P. \& {Lubow}, S.~H. 1994, The Astrophysical Journal, 421, 651

\bibitem[{{Artymowicz} \& {Lubow}(1996)}]{Artymowicz_Lubow_1996}
{Artymowicz}, P. \& {Lubow}, S.~H. 1996, \apjl, 467, L77

\bibitem[{{Bartos} {et~al.}(2017){Bartos}, {Kocsis}, {Haiman}, \&
  {M{\'a}rka}}]{Bartos_et_2017}
{Bartos}, I., {Kocsis}, B., {Haiman}, Z., \& {M{\'a}rka}, S. 2017, \apj, 835,
  165

\bibitem[{{Baruteau} {et~al.}(2011){Baruteau}, {Cuadra}, \&
  {Lin}}]{Baruteau_et_2011}
{Baruteau}, C., {Cuadra}, J., \& {Lin}, D.~N.~C. 2011, \apj, 726, 28

\bibitem[{{Belczynski} {et~al.}(2016){Belczynski}, {Holz}, {Bulik}, \&
  {O'Shaughnessy}}]{Belczynski_et_2016}
{Belczynski}, K., {Holz}, D.~E., {Bulik}, T., \& {O'Shaughnessy}, R. 2016,
  \nat, 534, 512

\bibitem[{{Breivik} {et~al.}(2016){Breivik}, {Rodriguez}, {Larson}, {Kalogera},
  \& {Rasio}}]{Breivik_et_2016}
{Breivik}, K., {Rodriguez}, C.~L., {Larson}, S.~L., {Kalogera}, V., \& {Rasio},
  F.~A. 2016, \apjl, 830, L18

\bibitem[{{Cuadra} {et~al.}(2009){Cuadra}, {Armitage}, {Alexander}, \&
  {Begelman}}]{Cuadra_et_2009}
{Cuadra}, J., {Armitage}, P.~J., {Alexander}, R.~D., \& {Begelman}, M.~C. 2009,
  \mnras, 393, 1423

\bibitem[{{Daddi} {et~al.}(2010){Daddi}, {Bournaud}, {Walter}, {Dannerbauer},
  {Carilli}, {Dickinson}, {Elbaz}, {Morrison}, {Riechers}, {Onodera}, {Salmi},
  {Krips}, \& {Stern}}]{Daddi_et_2010}
{Daddi}, E., {Bournaud}, F., {Walter}, F., {et~al.} 2010, \apj, 713, 686

\bibitem[{Deme {et~al.}(2020)Deme, Meiron, \& Kocsis}]{Deme2020}
Deme, B., Meiron, Y., \& Kocsis, B. 2020, The Astrophysical Journal, 892, 130

\bibitem[{{D'Orazio} {et~al.}(2013){D'Orazio}, {Haiman}, \&
  {MacFadyen}}]{DOrazio_et_2013}
{D'Orazio}, D.~J., {Haiman}, Z., \& {MacFadyen}, A. 2013, \mnras, 436, 2997

\bibitem[{{D'Orazio} \& {Samsing}(2018)}]{Dorazio_Samsing_2018}
{D'Orazio}, D.~J. \& {Samsing}, J. 2018, \mnras, 481, 4775

\bibitem[{{Duffell} {et~al.}(2019){Duffell}, {D'Orazio}, {Derdzinski},
  {Haiman}, {MacFadyen}, {Rosen}, \& {Zrake}}]{Duffell_et_2019}
{Duffell}, P.~C., {D'Orazio}, D., {Derdzinski}, A., {et~al.} 2019, arXiv
  e-prints, arXiv:1911.05506

\bibitem[{{Farris} {et~al.}(2014){Farris}, {Duffell}, {MacFadyen}, \&
  {Haiman}}]{Farris_et_2014}
{Farris}, B.~D., {Duffell}, P., {MacFadyen}, A.~I., \& {Haiman}, Z. 2014, \apj,
  783, 134

\bibitem[{{Fleming} \& {Quinn}(2017)}]{Fleming_Quinn_2017}
{Fleming}, D.~P. \& {Quinn}, T.~R. 2017, \mnras, 464, 3343

\bibitem[{{Ford} {et~al.}(2019){Ford}, {Bartos}, {McKernan}, {Haiman}, {Corsi},
  {Keivani}, {Marka}, {Perna}, {Graham}, {Ross}, {Stern}, {Bellovary}, {Berti},
  {O'Dowd}, {Lyra}, {MacLow}, \& {Marka}}]{Ford_et_2019}
{Ford}, K.~E.~S., {Bartos}, I., {McKernan}, B., {et~al.} 2019, \baas, 51, 247

\bibitem[{{Gr{\"o}bner} {et~al.}(2020){Gr{\"o}bner}, {Ishibashi}, {Tiwari},
  {Haney}, \& {Jetzer}}]{Paper_GWRates}
{Gr{\"o}bner}, M., {Ishibashi}, W., {Tiwari}, S., {Haney}, M., \& {Jetzer}, P.
  2020, arXiv e-prints, arXiv:2005.03571

\bibitem[{{Hayasaki}(2009)}]{Hayasaki_2009}
{Hayasaki}, K. 2009, \pasj, 61, 65

\bibitem[{{Hickox} {et~al.}(2014){Hickox}, {Mullaney}, {Alexander}, {Chen},
  {Civano}, {Goulding}, \& {Hainline}}]{Hickox_et_2014}
{Hickox}, R.~C., {Mullaney}, J.~R., {Alexander}, D.~M., {et~al.} 2014, \apj,
  782, 9

\bibitem[{{Hoang} {et~al.}(2018){Hoang}, {Naoz}, {Kocsis}, {Rasio}, \&
  {Dosopoulou}}]{Hoang_et_2018}
{Hoang}, B.-M., {Naoz}, S., {Kocsis}, B., {Rasio}, F.~A., \& {Dosopoulou}, F.
  2018, \apj, 856, 140

\bibitem[{{King} \& {Pounds}(2015)}]{King_Pounds_2015}
{King}, A. \& {Pounds}, K. 2015, \araa, 53, 115

\bibitem[{{King} {et~al.}(2007){King}, {Pringle}, \& {Livio}}]{King_et_2007}
{King}, A.~R., {Pringle}, J.~E., \& {Livio}, M. 2007, \mnras, 376, 1740

\bibitem[{{Kormendy} \& {Ho}(2013)}]{Kormendy_Ho_2013}
{Kormendy}, J. \& {Ho}, L.~C. 2013, \araa, 51, 511

\bibitem[{Le~Tiec(2015)}]{LeTiec2015}
Le~Tiec, A. 2015, Phys. Rev. D, 92, 084021

\bibitem[{{Levin}(2007)}]{Levin_2007}
{Levin}, Y. 2007, \mnras, 374, 515

\bibitem[{{Lubow} \& {Artymowicz}(2000)}]{Lubow_Artymowicz_2000}
{Lubow}, S.~H. \& {Artymowicz}, P. 2000, in Protostars and Planets IV, ed.
  V.~{Mannings}, A.~P. {Boss}, \& S.~S. {Russell}, 731

\bibitem[{{MacFadyen} \&
  {Milosavljevi{\'c}}(2008)}]{MacFadyen_Milosavljevic_2008}
{MacFadyen}, A.~I. \& {Milosavljevi{\'c}}, M. 2008, \apj, 672, 83

\bibitem[{{Mandel} \& {Farmer}(2018)}]{Mandel_Farmer_2018}
{Mandel}, I. \& {Farmer}, A. 2018, arXiv e-prints, arXiv:1806.05820

\bibitem[{{Mapelli}(2018)}]{Mapelli_2018}
{Mapelli}, M. 2018, arXiv e-prints, arXiv:1809.09130

\bibitem[{{Martin} {et~al.}(2019){Martin}, {Nixon}, {Pringle}, \&
  {Livio}}]{Martin_et_2019}
{Martin}, R.~G., {Nixon}, C.~J., {Pringle}, J.~E., \& {Livio}, M. 2019, \na,
  70, 7

\bibitem[{{Martini}(2004)}]{Martini_2004}
{Martini}, P. 2004, in Coevolution of Black Holes and Galaxies, ed. L.~C. {Ho},
  169

\bibitem[{{McKernan} {et~al.}(2018){McKernan}, {Ford}, {Bellovary}, {Leigh},
  {Haiman}, {Kocsis}, {Lyra}, {Mac Low}, {Metzger}, {O'Dowd}, {Endlich}, \&
  {Rosen}}]{McKernan_et_2018}
{McKernan}, B., {Ford}, K.~E.~S., {Bellovary}, J., {et~al.} 2018, \apj, 866, 66

\bibitem[{{Miranda} {et~al.}(2017){Miranda}, {Mu{\~n}oz}, \&
  {Lai}}]{Miranda_et_2017}
{Miranda}, R., {Mu{\~n}oz}, D.~J., \& {Lai}, D. 2017, \mnras, 466, 1170

\bibitem[{{Moody} {et~al.}(2019){Moody}, {Shi}, \& {Stone}}]{Moody_et_2019}
{Moody}, M. S.~L., {Shi}, J.-M., \& {Stone}, J.~M. 2019, \apj, 875, 66

\bibitem[{{M{\"o}sta} {et~al.}(2019){M{\"o}sta}, {Taam}, \&
  {Duffell}}]{Moesta_et_2019}
{M{\"o}sta}, P., {Taam}, R.~E., \& {Duffell}, P.~C. 2019, \apjl, 875, L21

\bibitem[{{Mu{\~n}oz} {et~al.}(2020){Mu{\~n}oz}, {Lai}, {Kratter}, \& {Mirand
  a}}]{Munoz_et_2020}
{Mu{\~n}oz}, D.~J., {Lai}, D., {Kratter}, K., \& {Mirand a}, R. 2020, \apj,
  889, 114

\bibitem[{{Mu{\~n}oz} {et~al.}(2019){Mu{\~n}oz}, {Miranda}, \&
  {Lai}}]{Munoz_et_2019}
{Mu{\~n}oz}, D.~J., {Miranda}, R., \& {Lai}, D. 2019, \apj, 871, 84

\bibitem[{{Murase} {et~al.}(2016){Murase}, {Kashiyama}, {M{\'e}sz{\'a}ros},
  {Shoemaker}, \& {Senno}}]{Murase_et_2016}
{Murase}, K., {Kashiyama}, K., {M{\'e}sz{\'a}ros}, P., {Shoemaker}, I., \&
  {Senno}, N. 2016, \apjl, 822, L9

\bibitem[{{Nishizawa} {et~al.}(2016){Nishizawa}, {Berti}, {Klein}, \&
  {Sesana}}]{Nishizawa_et_2016}
{Nishizawa}, A., {Berti}, E., {Klein}, A., \& {Sesana}, A. 2016, \prd, 94,
  064020

\bibitem[{{Perna} {et~al.}(2019){Perna}, {Lazzati}, \& {Farr}}]{Perna_et_2019}
{Perna}, R., {Lazzati}, D., \& {Farr}, W. 2019, \apj, 875, 49

\bibitem[{{Peters}(1964)}]{Peters_1964}
{Peters}, P.~C. 1964, Physical Review, 136, 1224

\bibitem[{{Pringle}(1981)}]{Pringle_1981}
{Pringle}, J.~E. 1981, \araa, 19, 137

\bibitem[{{Ragusa} {et~al.}(2016){Ragusa}, {Lodato}, \&
  {Price}}]{Ragusa_et_2016}
{Ragusa}, E., {Lodato}, G., \& {Price}, D.~J. 2016, \mnras, 460, 1243

\bibitem[{{Rodriguez} {et~al.}(2016){Rodriguez}, {Zevin}, {Pankow}, {Kalogera},
  \& {Rasio}}]{Rodriguez_et_2016}
{Rodriguez}, C.~L., {Zevin}, M., {Pankow}, C., {Kalogera}, V., \& {Rasio},
  F.~A. 2016, \apjl, 832, L2

\bibitem[{{Roedig} {et~al.}(2011){Roedig}, {Dotti}, {Sesana}, {Cuadra}, \&
  {Colpi}}]{Roedig_et_2011}
{Roedig}, C., {Dotti}, M., {Sesana}, A., {Cuadra}, J., \& {Colpi}, M. 2011,
  \mnras, 415, 3033

\bibitem[{{Shakura} \& {Sunyaev}(1973)}]{Shakura_Sunyaev_1973}
{Shakura}, N.~I. \& {Sunyaev}, R.~A. 1973, \aap, 500, 33

\bibitem[{{Shi} \& {Krolik}(2015)}]{Shi_Krolik_2015}
{Shi}, J.-M. \& {Krolik}, J.~H. 2015, \apj, 807, 131

\bibitem[{{Shi} {et~al.}(2012){Shi}, {Krolik}, {Lubow}, \&
  {Hawley}}]{Shi_et_2012}
{Shi}, J.-M., {Krolik}, J.~H., {Lubow}, S.~H., \& {Hawley}, J.~F. 2012, \apj,
  749, 118

\bibitem[{{Stone} {et~al.}(2017){Stone}, {Metzger}, \&
  {Haiman}}]{Stone_et_2017}
{Stone}, N.~C., {Metzger}, B.~D., \& {Haiman}, Z. 2017, \mnras, 464, 946

\bibitem[{{Tagawa} {et~al.}(2019){Tagawa}, {Haiman}, \&
  {Kocsis}}]{Tagawa_et_2019}
{Tagawa}, H., {Haiman}, Z., \& {Kocsis}, B. 2019, arXiv e-prints,
  arXiv:1912.08218

\bibitem[{{Tang} {et~al.}(2017){Tang}, {MacFadyen}, \& {Haiman}}]{Tang_et_2017}
{Tang}, Y., {MacFadyen}, A., \& {Haiman}, Z. 2017, \mnras, 469, 4258

\bibitem[{{Terquem} \& {Papaloizou}(2017)}]{Terquem_Papaloizou_2017}
{Terquem}, C. \& {Papaloizou}, J. C.~B. 2017, \mnras, 464, 2429

\bibitem[{{Thompson} {et~al.}(2005){Thompson}, {Quataert}, \&
  {Murray}}]{Thompson_et_2005}
{Thompson}, T.~A., {Quataert}, E., \& {Murray}, N. 2005, \apj, 630, 167

\end{thebibliography}

\end{document}